\documentclass{article}
\usepackage{graphicx}

\usepackage{graphpap}
\usepackage{amssymb,epsfig,amsmath}
\newcommand{\VB}{\begin{verbatim}}
\newcommand{\VE}{\end{verbatim}}

\newcommand{\de}{\,\mathrm d}

\newcommand{\Dar}[2]{\frac{\delta{#1}}{\delta{#2}}} 

\newcommand{\Par}[2]{\displaystyle\frac{\partial{#1}}{\partial{#2}}}

\newcommand{\ha}{\tfrac{1}{2}}

\newcommand{\tr}[1]{\mathrm{tr}(#1)}

\renewcommand{\(}{\left(}
\renewcommand{\)}{\right)}
\renewcommand{\[}{\left[}
\renewcommand{\]}{\right]}

\renewcommand{\b}[1]{{\bf{#1}}}

\newcommand{\nl}{\nonumber\\}

\begin{document}
\title{Three dimensional thermal-solute phase field  simulation of binary alloy solidification}
\author{P.C. Bollada\thanks{p.c.bollada@leeds.ac.uk},\and C.E. Goodyer\thanks{Now at: Numerical Algorithms Group: all other authors at University of Leeds U.K.},\and P.K. Jimack,\and A.M.Mullis, \and F.W. Yang }
\maketitle

\begin{abstract}
We employ adaptive mesh refinement,  implicit time stepping, a nonlinear multigrid solver  and  parallel computation, to solve a multi-scale, time dependent, three dimensional, nonlinear set of coupled partial differential equations for three scalar field variables. The mathematical model represents the non-isothermal solidification of a metal alloy into a melt substantially cooled below its freezing point at the microscale. Underlying physical molecular forces are captured at this scale by a specification of the energy field. The time rate of change of the temperature, alloy concentration and an order parameter to govern the state of the material (liquid or solid) is controlled by the diffusion parameters and variational derivatives of the energy functional. The physical problem is important to material scientists for the development of solid metal alloys and, hitherto, this fully coupled thermal problem has not been simulated in three dimensions, due to its computationally demanding nature. By bringing together state of the art numerical techniques this problem is now shown here to be tractable at appropriate resolution with relatively moderate computational resources.
\end{abstract}

\begin{figure}
\begin{center}
\includegraphics[width=15cm]{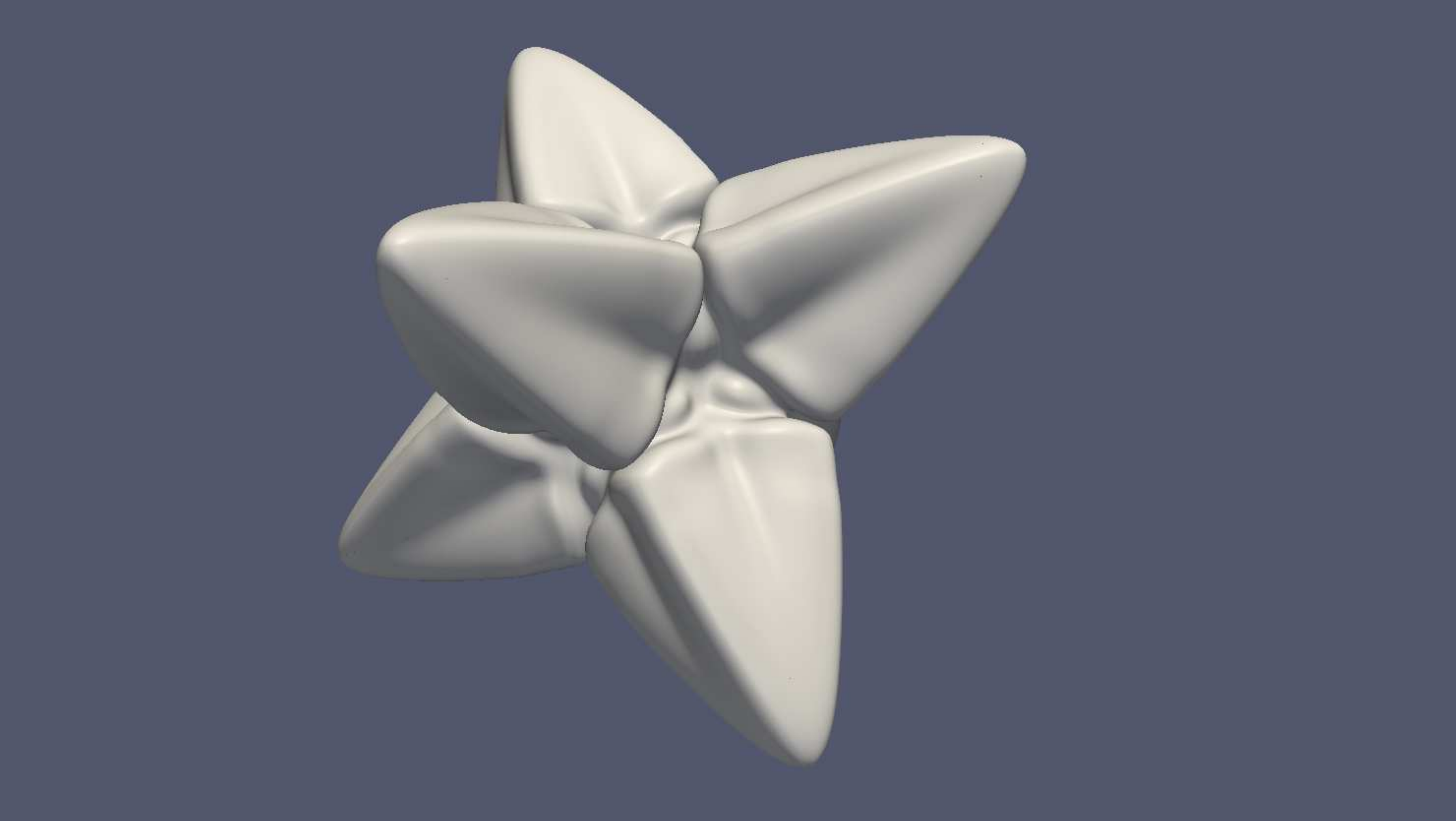}
\end{center}
\caption{Snapshot of the solid-liquid interface for a typical dendrite. This image was obtained from a simulation with $Le=40,\Delta=0.525$ and $\Delta x=0.78$.}
\label{m9}
\end{figure}

\section{Introduction}

We here present our computational approach to simulating, at the meso scale, three dimensional, non-isothermal, alloy solidification   from an initial small, spherical seed into a mature, dendritic crystal. A feature of a mature dendrite is the geometric complexity of its evolving two-dimensional surface (see Fig. \ref{m9} for a typical snapshot in time). This makes tracking of the surface a difficult task in sharp interface models.  A phase-field model avoids this by making  use of the phase field, $\phi({\bf x},t)\in[-1,1]$, to represent, at its two extremes, the liquid and solid state respectively and  the evolution of the phase boundary,  $\phi=0$, is the surface of  interest.  This solves one problem, but at a cost of introducing another. The computation requires an extra variable, the phase, which varies rapidly over a small region about the interface.  Taking the thickness of the interface to be  $\approx 1$, we find the size of a mature dendrite grows to $\sim 300$, requiring the domain size to be significantly greater still (depending on the thermal field this may need to be $O(1000)$, or even more), we see that the interface region of interest is very much smaller than the overall domain. Of major concern in phase field models is the dependence of the computed results on the interface width. To address this, Karma \cite{karma2001}, analysed the problem in the thin interface limit to produce a phase field formulation that is independent of the interface width up to several orders of magnitude larger than a physically real value, although this is less clear when a thermal field is coupled. That said, we adopt an  interface width that is of physically realistic order. There are two coupled driving forces for growth: the alloy concentration, governed by a diffusion parameter $D_c$ and, secondly, a temperature field governed by a diffusion parameter $D_\theta$. The ratio of these two parameters gives the Lewis number, $Le=D_\theta/D_c$, which for many metallic alloys approaches $10,000$. In two dimensions  Lewis numbers of this magnitude have been realised by \cite{Rosam2009}. However, there are are no prior results, even for the interface widths permitted by Karma's model, for even very moderate Lewis number in three dimensions. This paper seeks to demonstrate that such results are feasible provided the appropriate numerical techniques are employed.

The numerical solution to this phase field  model (described in detail in the following section) requires methods to solve a time-dependent, highly nonlinear system of PDEs, of parabolic type, and capable of resolving varying length and time scales. A feature that adds another level of difficulty to this problem is that we are particularly interested in the tip radius and speed of growth of the dendrite only when it is fully mature, and these two numbers are steady.  In summary, the computational problem is: non-linear, three-dimensional, stiff, involves multiple length scales to capture small phase  and large temperature fields, multi-time scale associated with the Lewis number and, to establish   a mature dendrite, requires a long simulation time.

The computational techniques we employ are: use of very fine meshing in the region around the moving boundary where phase field and solute field resolution is critical, and coarse meshing away from the boundary where only the slowly changing temperature field requires  resolution; implicit time stepping to allow much larger time steps than would otherwise be possible;  nonlinear smoothing in conjunction with a nonlinear multi-grid solver; and parallel processing with up to 1024 cores as the simulation progresses. The combination of all of these techniques allows an almost optimal solution process to be developed, in which the number of degrees of freedom is evolved with the dendrite, to maintain the required resolution as the interface grows, and the solution time at each time step is approximately proportional to the number of degrees of freedom. Furthermore, the use of a parallel implementation ensures that sufficient primary memory is available to support a mesh resolution which is fully converged whilst maintaining a tractable solution time.

The particular phase field model we employ is an extension of \cite{citeulike:12734814}, and is based on the three dimensional thermal- phase field model of \cite{Karma199754} and two dimensional thermal-solutal phase field model of  \cite{citeulike:12734676}.  One feature of the physical problem is that it is purely dissipative, or entropy increasing, as all natural relaxational phenomena are. The resulting PDEs are of Allen-Cahn \cite{Allen1979} and Carn-Hilliard type \cite{Cahn1961}. That is to say,  the model involves time derivatives of the three fields coupled to forms involving variational derivatives of some functional - typically the free energy functional. As the dendrite grows the free energy reduces monotonically with time but never achieves equilibrium if the domain boundary is far from the dendrite. Although we have listed some of the difficult aspects of this model, the relaxational aspect is typically an asset and results in stable numerical schemes: there is no convection, for example (at least in the absence of flow in the melt).

 The variational form of the mathematical model is, of course,  identical to the two-dimensional model in form. However, on realising the variational derivatives the resulting equations are more complex and nonlinear in the higher dimension. This is largely because of surface energy related  anisotropy  associated to alignment at the molecular scale.  In two dimensions anisotropy is conveniently formulated using a single angle parameter, but in three dimensions we prefer to use a normal given in terms of Cartesian gradients.

In addition to the Lewis number,  another key parameter in the simulations is the undercooling, $\Delta$, which sets the temperature of the liquid's initial and far boundary condition below its freezing point. As this parameter becomes larger the under-cooling becomes more severe, the solification more rapid and fractal in appearance: and, also, correspondingly more difficult to simulate.

The other field, not hitherto discussed, is the solute field. For a binary alloy of initial concentration, $c_\infty$, the concentration of an alloy component at any point is represented by a value of $c({\bf x},t)\in[0,1]$. The requirement for equilibrium at the solid-liquid interface means that the concentration in the solid and the concentration in the liquid at the interface will be unequal. In a sharp interface model this results in a discontinuous jump in $c$ at the interface, while in phase field models it results in a steep, but continuous, increase in c across the diffuse interface region, where there is some advantage in reformulating the problem to remove this. We show this in the next section.

\section{Governing equations}
The governing equations for dendritic growth of an under-cooled binary alloy are here presented in full, in both their variational form and in the (equivalent) form of PDEs for numerical implementation. The non-dimensional equations for the phase field, $\phi$, the solute concentration, $c$ and the dimensionless temperature, $\theta$, are given via a specification of the free energy
\begin{align}
F \equiv \int_V \ha A(\b n)^2\nabla\phi\cdot\nabla\phi+f(\theta,\phi)\de V
\label{F}
\end{align}
and the relations
\begin{subequations}
\begin{align}
\label{Sa}
\tau(c,\phi)A^2(\b n)\dot{\phi} &= -\Dar{F}{\phi}\\
\label{Sb}
\dot c &= \nabla\cdot\left(K(\phi)\nabla\Dar{F}{c}-\b j\right),\\
\label{Sys}
\dot \theta &= D_\theta\nabla^2 \theta+\ha\dot\phi.
\end{align}
\end{subequations}
The solute diffusion parameter is given by
\begin{align}
K = D_c\ha(1-\phi).
\end{align}
The parameter $D_c$   is a diffusion constant, thus $K=0$ in the solid ($\phi=1$) and $K=D_c$ in the liquid ($\phi=-1$). $D_\theta$ is the temperature diffusion coefficient (assumed constant). The normal to the inteface is given by
\begin{align}
\b n=\frac{\nabla\phi}{|\nabla\phi|},
\end{align}
which is well defined around $\phi=0$, and the anisotropy function for cubic symmetry (growth is preferred along the normals to the faces)  is given for three dimensions by \cite{Karma199754},
\begin{align}
A(\b n)&\equiv A_0\[1+\tilde\epsilon\(n_x^4+n_y^4+n_z^4\)\]
\label{A}
\end{align}
where ${\b n}=[n_x,n_y,n_z]$, $A_0=1-3\epsilon$, $\tilde{\epsilon}=4\epsilon/(1-3\epsilon)$ and $\epsilon\approx 0.02$ governs the amount of anisotropy. The reason for this arrangement of constants is to compare with the two dimensional form
\begin{align}
A(\b n)\equiv  A_0\[1+\tilde\epsilon\(n_x^4+n_y^4\)\]\equiv 1+\epsilon\cos 4\psi
\end{align}
where the angle, $\psi$, is given by 
\begin{align}
\tan\psi=\Par{\phi}{y}/\Par{\phi}{x}.
\end{align}
The dimensionless relaxation time function is defined by
\begin{align}
\tau(c,\phi)\equiv \frac{1}{Le}+Mc_\infty[1+(1-k_E)U],
\label{tau}
\end{align}
where the Lewis number $Le= D_\theta/D_c$ and
\begin{align}
U\equiv \frac{1}{1-k_E}\left(\frac{2c/c_\infty}{1+k_E-(1-k_E)\phi}-1\right).
\end{align}
Here $k_E$ is the equilibrium partition coefficient, $c_\infty$ is the far boundary condition for $c$. The anti-trapping current $\b j$, appearing in the solute equation, Eq. \ref{Sb}, is prescribed by
\begin{align}
\b j = - \frac{1}{2\sqrt 2}[1+(1-k_E)]U\dot\phi\b n,
\end{align}
The profile of $c$ exhibits a spike at the interface, which can present computational difficulties. Following \cite{citeulike:12734676}, this is largely overcome by rewriting the solute equation using the variable $U$: 
\begin{eqnarray}
\left(\frac{1+k_E}{2}-\frac{1-k_E}{2}\phi\right)\Par{U}{t} = \nabla\cdot\left\{D_c\frac{1-\phi}{2}\nabla U+{\b j}\right\}+\frac{1}{2}[1+(1-k_E)U]\Par{\phi}{t}.
\label{Ueqn}
\end{eqnarray}
The physical temperature field, $T$, can be recovered by the relationship
\begin{align}
\theta=\frac{T-T_M-mc_\infty}{L/C_p},
\end{align}
where  $L$ and $C_p$ are the latent heat of the phase transition and heat capacity respectively. The slope of the liquidus line is given by $m = M L/[C_p(1-\kappa_E)]$ and $T_M$ is the melting temperature  of the alloy. 

Finally the bulk free energy density is given by
\begin{align}
f(\theta,\phi)\equiv \frac{\phi^2}{2}\left(\frac{\phi^2}{2}-1\right)+\lambda(\theta+c_\infty U)\left(\phi-\frac{2\phi^3}{3}+\frac{\phi^5}{5}\right).
\label{f}
\end{align}
We solve the  system of equations \ref{Sa},\ref{Sys} and \ref{Ueqn} plus initial (typically small) solid seed see subsection \ref{boundary}  and far boundary conditions 
\begin{align}
\phi|_\text{far}&=-1\nl
 U|_\text{far}&=0\quad (\equiv c|_\text{far}= c_\infty)\nl
\theta|_\text{far}&= -\Delta
\end{align}
where $\Delta$ is the given under-cooling. The equation for temperature is a standard diffusion equation \footnote{without convection due to no velocity field} with a heating term, $\dot \phi$, proportional to the  solidification rate  (or cooling if melting).  The driving force for the phase equations is given by  $f(T,\phi)$, consisting of a  double well potential having stable minima at $\phi=\pm 1$ and a maximum at $\phi=0$ and a function of $\theta$ to create conditions for moving the phase boundary. For example a negative value of $\theta$ creates conditions favourable for solidification. The parameter, $\lambda$, is proportional to the interface width, which in turn is chosen as the characteristic length scale.

\subsection{Parameter values}
For the purposes of this paper we choose a selection of parameters to use as default values for the simulations below in Table  \ref{T}. Any deviation from these parameter values is explicitly noted in the text. 
\begin{table}[h]
\begin{tabular}{|c|c|l|}
\hline
\hline
Physical property& Symbol& value\\
\hline
Anisotropy&$\epsilon$&0.02\\
Boundary concentration& $Mc_\infty$&0.05\\
Equilibrium partition coefficient&$\kappa_E$&0.3\\
Dimensionless interface width&$\lambda$&2\\
Ratio of solute diffusivity to characteristic diffusivity&$D_c$&1.2534\\
Lewis number - $D_\theta/D_c$&Le&40 and 100\\
Dimensionless Undercooling at the far boundary&$\Delta$&0.25 to 0.80\\
Initial nuclear radius& $R_0$&5.0\\
\hline
\hline
Computational property&symbol&value\\
\hline
Finest grid size&$\Delta x$& 0.195 to 0.78\\
Computation domain size&$L^3$&800$\times$800$\times$800\\
\hline
\end{tabular}
\caption{Table of parameter values used for the simulations in this paper.}
\label{T}
\end{table}

\subsection{Anisotropic calculations}
Note that the phase equation, Eq. \ref{Sa}, is made considerably more complicated by the presence of the anisotropy term, $A(\b n)$, in the free energy functional. The variational derivative of a functional not involving gradients is simply the partial derivative of the density with respect to that variable. Thus
\begin{align}
\Dar{}{\phi}\int f(T,\phi)\de V &=\Par{f}{\phi}\nl
&=\phi^3-\phi+\lambda(\theta+c_\infty U)(1-2\phi^2+\phi^4).
\label{f_phi}
\end{align}
The variational derivative of the pure gradient part of the functional is given by
\begin{align}
\Dar{G}{\phi}\equiv\Dar{}{\phi}\int g(\nabla\phi)\de V= -\nabla\cdot\left(\left.\Par{}{\bf r}g({\bf r})\right|_{\bf r=\nabla\phi}\right)
\label{div}
\end{align}
where, in our model,
\begin{align}
 g({\bf r})\equiv\ha A(\b n)^2|{\bf r}|^2, \text{ for } {\bf r}\in \mathbb{R}^3
 \label{G}
\end{align}
In order to expand Eq. \ref{div} and thus, Eq. \ref{Sa}, we first introduce the notation: $\phi_{,i}\equiv \partial_i\phi\equiv\Par{\phi}{x^i}$ etc., for Cartesian differentiation, and subscripts for differentiation on function space. Thus
\begin{align}
 g_{i}&\equiv \Par{g}{r_i},\nl
 g_{ij}&\equiv \Par{}{r_i}\Par{g}{r_j}.
\end{align}
This enables us to write
\begin{align*}
 -\Dar{G}{\phi}&= \partial_ig_i \nl
  &= \phi_{,ij} g_{ij}\quad\text {(using the chain rule)},
\end{align*}
which written out in full reads
\begin{align*}
 -\Dar{G}{\phi}\equiv-\Dar{}{\phi}\int g(\nabla\phi)\de x^3=\Par{{}^2\phi}{x^i\partial x^j}\left.\left(\Par{}{r_i}\Par{g}{r_j}\right)\right|_{\bf r=\nabla \phi}.
\end{align*}
Note that, $g_{ij}$, is a function of only first derivatives of $\phi$. To avoid expanding the above in terms of the components, $\phi_{,i}$ we first introduce  the substitutions $q\equiv |\nabla\phi|^2$  and  ${\bf X}\equiv[X_1,X_2,X_3]\equiv[\phi_{,1}^2/q,\phi_{,2}^2/q,\phi_{,3}^2/q]$, to write the anisotropy 
\begin{equation}
A=A_0\left(1+\tilde\epsilon\sum_{i=1}^3 X^2_i\right).
\end{equation}
For an arbitrary function $h({\bf r})=\tilde h({\bf r},q,{\bf X}, A)$ we use the chain rule to write
\begin{eqnarray}
\Par{h}{r_i}&=&\(\Par{}{r_i}+\Par{q}{r_i}\Par{}{q}+\Par{X_j}{r_i}\Par{}{X_j}+\Par{A}{r_i}\Par{}{A}\)\tilde h\nl
&=&\(\Par{}{r_i}+2r_i\Par{}{q}+\frac{2r_i}{q}(\delta_{ij}-X_j)\Par{}{X_j}+\frac{2r_i}{q}[2A_0\tilde\epsilon X_i-2(A-A_0)]\Par{}{A}\)\tilde h,
\end{eqnarray}
where we have used
\begin{align}
\Par{q}{r_i}&=2r_i,\nl
\Par{X_j}{r_i}&\equiv \Par{}{r_i}\left(\frac{r_j^2}{q}\right)=\frac{2r_i\delta_{ij}}{q}-\frac{2r_ir_j^2}{q^2}=\frac{2r_i}{q}(\delta_{ij}-X_j),\nl
\Par{A}{r_i}&=\Par{X_j}{r_i}\Par{A}{X_j}=\frac{2r_i}{q}(\delta_{ij}-X_j)\Par{A}{X_j}\nl
&=\frac{2r_i}{q}(\delta_{ij}-X_j)2A_0\tilde\epsilon X_j\nl
&=\frac{4r_i}{q}(A_0\tilde\epsilon X_i-A+A_0)
\end{align}
using, on the last simplification, the identity $A-A_0\equiv A_0\tilde\epsilon \sum_j X_j^2$.  This allows us to  compute
\begin{align}
g_i\equiv \Par{}{r_i}\(\ha A^2q\)|_{\bf r=\nabla\phi}
&=\phi_{,i}A^2+4\phi_{,i}(A_0X_i\tilde\epsilon-A+A_0)A
\end{align}
and further differentiation gives the concise forms
\begin{eqnarray}
g_{ii}&=&(24X_i-3)A^2+(-48X_i^2\tilde\epsilon+12X_i\tilde\epsilon-40X_i+4)
A_0A+16X_i(X_i\tilde\epsilon+1)^2A_0^2\nl
g_{ij}&=&\frac{\phi_{,i}\phi_{,j}}{g}\left[24A^2+(-24X_i\tilde\epsilon-24X_j\tilde\epsilon-40)A_0A+(16(X_j\tilde\epsilon+1))(X_i\tilde\epsilon+1)A_0^2\right],\quad i\ne j.\nl
\label{gij}
\end{eqnarray}
The above expressions are not only much more concise than the expanded equivalent as a function of $\phi_{,i}$, but are functions of $X_i$ and $A$ which are of order unity in size and thus minimise floating point errors (the expanded equivalent contains tenth order polynomials of $\phi_{,i}$).
We note also that, in the absence of anisotropy,  $\tilde\epsilon=0$, reduces $g_{ij}$ to $\delta_{ij}$ so that $\phi_{,ij}g_{ij}=\nabla^2\phi$. In the situation where $g_{ij}$ is ill defined due to $|\nabla\phi|\rightarrow 0$ we set.
\begin{equation}
\phi_{,ij}g_{ij}|_{|\nabla\phi|\rightarrow 0}=A_0^2(1+\tilde{\epsilon})^2\nabla^2\phi
\end{equation}
or, equivalently $A|_{|\nabla\phi|\rightarrow 0}=A_0(1+\tilde{\epsilon})$. In practice we use this expression only when $|\nabla\phi|=0$, to machine precision without difficulty.

Using the notation  $\tr{{\bf g}}\equiv \delta_{ij}g_{ij}$, the rearrangement
\begin{eqnarray}
\phi_{,ij}g_{ij}&\equiv&\tfrac{1}{3}(\phi_{,11}+\phi_{,22}+\phi_{,33})(g_{11}+g_{22}+g_{33})+(\phi_{,ij}-\tfrac{1}{3}\phi_{,kk}\delta_{ij})g_{ij}\nl
&\equiv&\tfrac{1}{3}\nabla^2\phi\, \text{tr}({\bf g})+(\phi_{,ij}-\tfrac{1}{3}\phi_{,kk}\delta_{ij})g_{ij},
\label{R01}
\end{eqnarray}
allows the dominant term to be isolated and has advantage, because the Laplacian can be discretised to minimise grid induced anisotropy. We will also see that the term $(\phi_{,ij}-\tfrac{1}{3}\phi_{,kk}\delta_{ij})$, like $g_{ij}$, on discretisation with a compact stencil at a discrete node, ${\bf p}$,  only has contributions from the nodes surrounding ${\bf p}$. This affords simplification for the non-linear solver later discussed.
\subsection{System summary}
 Writing, $M_{ij}\equiv \phi_{,ij}-\tfrac{1}{3}\phi_{,kk}\delta_{ij}$ we summarise  the nonlinear PDE system that forms our mathematical model as
\begin{align}
\tau(c,\phi)A(\b n)^2\dot\phi=\tfrac{1}{3}\nabla^2\phi\, \text{tr}({\bf g})+M_{ij}g_{ij}-\Par{f}{\phi}
\label{phiequation}
\end{align}
where $\tau(c,\phi)$ is given by Eq. \ref{tau}, $\Par{f}{\phi}$ is given in Eq. \ref{f_phi}, $g_{ij}$ in Eq. \ref{gij}, the solute is solved via Eq. \ref{Ueqn} and the temperature by Eq. \ref{Sys}.

\section{Discretisation}
The approach taken to discretisation is based upon a cell centred finite difference scheme, in that the nodes of the domain are located at the centre of cubic cells. and thus, we use the term `node' and `cell centre' inter changeably. One consequence of this is that there are no nodes on the domain boundary, thus   making the use of Dirichlet boundary conditions non-trivial. The scheme makes use of the PARAMESH library to support mesh adaptivity in parallel \cite{MacNeice2000330,olson2006}. The meshes obtained by this approach take the form of an oct tree of regular blocks, within which the mesh is uniform, and it is the spatial discretisation on any one of these blocks that we discuss here. Subsequently we will discuss   adaptive mesh refinement and  the implicit temporal discretisation scheme that is employed.

\subsection{Spatial discretisation}
Compact finite difference stencils $(3\times 3\times3)$, are used to discretise the first and second derivatives.  Denoting these 27 points by ${\bf Q}$ and defining a generic 27 point Laplacian stencil, $W_{abc}$, around a point ${\bf p}=[i,j,k]$  by
\begin{eqnarray}
\nabla^2 	u|_{\bf Q}&=&\frac{1}{(\Delta x)^2}\sum_{a=-1}^1\sum_{b=-1}^1\sum_{c=-1}^1 W_{abc}u_{{\bf p}+[a,b,c]}
\label{Lap}
\end{eqnarray}
where   $\Delta x$ is the physical distance between nearest neighbours, we can recover the 7 point Laplacian stencil, built from only the centre node, ${\bf p}$ and the 6 nearest neighbours ($a^2+b^2+c^2=1$)
\begin{eqnarray}
\nabla^2u|_{\bf Q}=\frac{-6 u_{i,j,k}+u_{i+1,j,k}+u_{i-1,j,k}+u_{i,j+1,k}+u_{i,j-1,k}+u_{i,j,k+1}+u_{i,j,k-1}}{(\Delta x)^2}
\end{eqnarray}
by setting the weights
\begin{eqnarray}
W_{abc}=\left\{\begin{array}{ll}
-6&a^2+b^2+c^2=0\\
1&a^2+b^2+c^2=1\\
0&\text{otherwise}
\end{array}\right..
\end{eqnarray}
However, this stencil is more prone to  grid anisotropy than the following 27 point Laplacian stencil  (see \cite{SpotxCarey1996}), with weights
\begin{eqnarray}
W_{abc}=\left\{\begin{array}{ll}
-128/30&a^2+b^2+c^2=0\\
14/30&a^2+b^2+c^2=1\\
3/30&a^2+b^2+c^2=2\\
1/30&a^2+b^2+c^2=3
\end{array}\right.
\label{W27}
\end{eqnarray}

In order to discretise the phase equation, Eq.\ref{phiequation}, in space it is necessary to approximate $\phi_{,ij}g_{ij}$ about the point ${\bf p}$. Using the above notation we obtain:
\begin{equation}
(\phi_{,ij}g_{ij})|_{{\bf Q}}=\tfrac{1}{3}\nabla^2\phi|_{{\bf Q}}\, \tr{{\bf g}}|_{({\bf Q-p})}+M_{ij}|_{({\bf Q-p})}g_{ij}|_{({\bf Q-p})},
\label{D01}
\end{equation}
where we use the notation, ${}|_{{\bf Q-p}}$,  to denote that the central node is not used.
We discretise, $M_{ij}$ as follows
\begin{eqnarray}
\Delta|_{\bf Q-p}&\equiv& \phi_{{\bf p}+[1,0,0]}+\phi_{{\bf p}+[-1,0,0]}+\phi_{{\bf p}+[0,1,0]}+\phi_{{\bf p}+[0,-1,0]}+\phi_{{\bf p}+[0,0,1]}+\phi_{{\bf p}+[0,0,-1]}\nl
M_{11}|_{{\bf Q -p}}&=&\tfrac{1}{\Delta x^2}\left(\phi_{{\bf p}+[1,0,0]}+\phi_{{\bf p}+[-1,0,0]}-\tfrac{1}{3}\Delta|_{\bf Q-p}\right),\nl
M_{22}|_{{\bf Q -p}}&=&\tfrac{1}{\Delta x^2}\left(\phi_{{\bf p}+[0,1,0]}+\phi_{{\bf p}+[0,-1,0]}-\tfrac{1}{3}\Delta|_{\bf Q-p}\right),\nl
M_{33}|_{{\bf Q -p}}&=&\tfrac{1}{\Delta x^2}\left(\phi_{{\bf p}+[0,0,1]}+\phi_{{\bf p}+[0,0,-1]}-\tfrac{1}{3}\Delta|_{\bf Q-p}\right),\nl
M_{12}|_{{\bf Q -p}}=M_{21}|_{{\bf Q -p}}&=&\tfrac{1}{4\Delta x^2}\left(\phi_{{\bf p}+[1,1,0]}+\phi_{{\bf p}+[-1,-1,0]}-\phi_{{\bf p}+[1,-1,0]}-\phi_{{\bf p}+[-1,1,0]}\right)\nl
M_{23}|_{{\bf Q -p}}=M_{32}|_{{\bf Q -p}}&=&\tfrac{1}{4\Delta x^2}\left(\phi_{{\bf p}+[0,1,1]}+\phi_{{\bf p}+[0,-1,-1]}-\phi_{{\bf p}+[0,-1,1]}-\phi_{{\bf p}+[0,1,-1]}\right)\nl
M_{31}|_{{\bf Q -p}}=M_{13}|_{{\bf Q -p}}&=&\tfrac{1}{4\Delta x^2}\left(\phi_{{\bf p}+[1,0,1]}+\phi_{{\bf p}+[-1,0,-1]}-\phi_{{\bf p}+[1,0,-1]}-\phi_{{\bf p}+[-1,0,1]}\right),
\label{Mij}
\end{eqnarray}
where ${}|_{\bf Q -p}$ denotes use of some or all of the $3^3-1$ surrounding nodes, ${\bf p}+[1,0,0],{\bf p}+[0,1,0],...,{\bf p}+[1,1,1]$. The matrix elements, $g_{ij}$, are functions of the components of $\nabla\phi$ only:
\begin{align}
\phi_{,1}|_{\bf Q-p}&=\tfrac{1}{2\Delta x}\left(\phi_{{\bf p}+[1,0,0]}-\phi_{{\bf p}+[-1,0,0]}\right)\nl
\phi_{,2}|_{\bf Q-p}&=\tfrac{1}{2\Delta x}\left(\phi_{{\bf p}+[0,1,0]}-\phi_{{\bf p}+[0,-1,0]}\right)\nl
\phi_{,3}|_{\bf Q-p}&=\tfrac{1}{2\Delta x}\left(\phi_{{\bf p}+[0,0,1]}-\phi_{{\bf p}+[0,0,-1]}\right)
\label{FirstOrder}
\end{align}
Consequently, Eq. \ref{D01} has the property that only $\nabla^2\phi|_{{\bf Q}}$ contains a contribution from the central node, $\phi_{\bf p}$ and thus
\begin{equation}
\Par{}{\phi_{\bf p}}(\phi_{,ij}g_{ij})|_{\bf Q}=\tfrac{1}{3}\tr{{\bf g}}|_{\bf Q-p}\Par{}{\phi_{\bf p}}\nabla^2\phi|_{\bf Q}=-\frac{128}{90}\, \tr{{\bf g}}|_{{\bf Q-p}}.
\end{equation}
This is important  for the Jacobi linearisation described in the next section. The PDE for $\phi$ is thus approximated by ODEs at each point, ${\bf p}$, by
\begin{align}
\dot \phi_{{\bf p}} = F^\phi_{{\bf p}}(\phi_{{\bf Q}},U_{{\bf p}},\theta_{{\bf p}})
\end{align}
where
\begin{align}
 F^\phi_{{\bf p}}&\equiv\frac{\tfrac{1}{3}\nabla^2\phi|_{{\bf Q}}\tr{{\bf g}}|_{{\bf Q-p}}+M_{ij}|_{{\bf Q-p}}\,g_{ij}|_{{\bf Q-p}}-\Par{f}{\phi}(\phi_{{\bf p}},U_{{\bf p}},\theta_{{\bf p}})}{\tau(c_{{\bf p}},\phi_{{\bf p}})A^2|_{{\bf Q-p}}}.
 \label{phirhs}
\end{align}
In the above $\tau(c_{{\bf p}},\phi_{{\bf p}})$ is given by Eq. \ref{tau}  and $\Par{f}{\phi}|_{{\bf p}}$ is given by Eq. \ref{f_phi} using the values for $\phi,U,\theta$ at point ${\bf p}$. The Laplacian $ \nabla^2\phi|_{{\bf Q}}$ is given by Eq. \ref{Lap} with weights Eq. \ref{W27}. The indexed functions $M_{ij}$ are given by Eq. \ref{Mij} and the functions $g_{ij}$ are given by Eq. \ref{gij}, where $A,X_i,|\nabla\phi|^2$ are all functions of $\phi_{,i}$ approximated by second order differences, Eq. \ref{FirstOrder}. The PDE for $U$, Eq	. \ref{Ueqn}, is approximated by the ODEs
\begin{align}
\dot{U}_{{\bf p}}=F^U_{{\bf p}}(\dot\phi_{{\bf p}},\phi_{{\bf p}},U_{{\bf Q}},\theta_{{\bf p}})\nl
\end{align}
where
\begin{align}
F^U_{{\bf p}}&\equiv  \frac{\nabla\cdot\left\{D_c\frac{1-\phi_{\bf p}}{2}\nabla U|_{\bf Q}+{\b j}\right\}+\frac{1}{2}[1+(1-k_E)U_{\bf p}]\dot\phi_{\bf p}}{\left(\frac{1+k_E}{2}-\frac{1-k_E}{2}\phi_{\bf p}\right)}
\end{align}
which is expanded in full (shown in subsection \ref{temporal}) with the same derivative discretisation scheme used for $U$ as $\phi$.

Finally the $\theta$ term is given by
\begin{align}
\dot\theta_{\bf p} = F^\theta_{\bf p}(\theta_{\bf Q},\dot\phi_{\bf p})\equiv D_\theta\nabla^2\theta|_{\bf Q}+\ha\dot\phi_{\bf p}.
\end{align}

\subsection{Boundary and initial conditions}
\label{boundary}
We use zero Neumann boundary conditions for all variables. This is easily implemented by imposing values to the ghost cells of all  blocks adjacent to the Neumman boundary, that are equal to the cell values of those cells next to the boundary (see below for more discussion of ghost cells). In the discretisation, this sets all boundary derivatives equal to zero. In exploitation of the symmetry in the problem this is interpreted as reflective symmetry on the planes $x=0,y=0,z=0$ and, provided the domain is sufficiently  large, as equivalent to  Dirichlet conditions on the far boundary for all variables. If, during a simulation, the normal derivative of any of the dependent variables begins to deviate from zero by more than a prescribed tolerance then we may allow the domain to expand so as to ensure we retain a zero normal derivative on the revised boundary (see below for more details of the mesh adaptivity that facilitates this).For a fixed domain, it is important that the dimensions are sufficient not only to represent the growing dendrite but also the temperature field throughout the simulation, which typically extends well ahead of the phase interface (especially for large values of the Lewis number).

The initial conditions for this problem are to some extent flexible as the evolution of the variables in time will alter physically inappropriate starting conditions. Thus the initial phase profile, temperature and concentration field, in general, will all adjust in the very early stages of the simulation. The temperature field can take longer to adjust to a profile which is near the melting point of the alloy inside the solid if started at a constant field value and, so we anticipate this with the condition given below.

The initial condition for the phase field with seed radius given by $R$ is prescribed by
\begin{align}
\phi(t=0,{\bf x(p)})=-\tanh [\alpha(\sqrt{{\bf x\cdot x}}-R)],
\end{align}
where we employ the factor $\alpha=0.6$, the precise value of which is not important as the solver smooths the phase profile if $\alpha$ is large and conversely sharpen the profile if $\alpha$ is too small within reason. It is not found necessary to normalise this profile so that $\phi(t=0,{\bf x=0})=1$. The initial solute condition is $U=0$ and the temperature profile used is
\begin{align}
\theta(t=0,{\bf x(p)})= -\Delta +\ha\Delta(\phi+1).
\end{align}
The most significant parameter in the initial conditions, in terms of the sensitivity of the subsequent calculations, is the radius of the initial nucleus. It has been shown that the transient behaviour of the evolving dendrite can be affected by this value well into the simulation,  \cite{Rosam20084559} (though the final geometry and velocity of the dendrite tip is much less sensitive). To this end we choose the smallest value of $R$ such that the dendrite does not melt (melting can occur if there is   insufficient solid $\phi=1$ in the nucleus due the encroachment of the diffuse interface near the nucleus centre). We find the smallest value to be $R\approx 5$. 
\subsection{Temporal discretisation}
\label{temporal}
Due to the stiffness of the nonlinear system of ODEs that arises following the spatial discretization we employ BDF2 time stepping, so that  at a point, ${\bf p}$ in the grid domain at the centre the $3^3$ points, ${\bf Q}$, the phase field variable system is approximated by
\begin{align}
\phi_{{\bf p}}^{n+1}-r_2\phi_{{\bf p}}^{n}+r_3\phi_{{\bf p}}^{n-1}=r_1\Delta t^{n+1} F^\phi_{{\bf p}}
\label{C_time}
\end{align}
In practice, we introduce another variable
\begin{align}
\phi^*_{{\bf p}}\equiv r_2\phi_{{\bf p}}^{n}-r_3\phi_{{\bf p}}^{n-1}
\end{align}
and write Eq. \ref{C_time} as
\begin{align}
\phi_{{\bf p}}^{n+1}-\phi^*_{{\bf p}}=r_1\Delta t^{n+1}  F^\phi_{{\bf p}}(\phi_{{\bf Q}}^{n+1},U_{{\bf p}}^{n+1},\theta_{{\bf p}}^{n+1}).
\label{phi_time}
\end{align}
The right hand side, $F^\phi_{{\bf p}}$,  is defined by Eq. \ref{phirhs}.

For constant time step, $r_1=2/3,r_2=4/3,r_3=1/3$. For a growing dendrite it is essential to use a small time step at the initial (imposed) state. Thereafter the time step is increased (see subsection \ref{Nonlinear} for detail) to fully exploit the implicit time stepping. The BDF2 for  adaptive time stepping is given by
\begin{align}
r_1&\equiv(r+1)/(2r+1),\nl
r_2&\equiv(r+1)^2/(2r+1),\nl
r_3&\equiv r^2/(2r+1)\nl
r&\equiv \Delta t^{n+1}/\Delta t^n
\end{align}
where $r$ is the ratio of the current to previous time step.

Similarly, with $U^*_{{\bf p}}\equiv r_2U_{{\bf p}}^{n}-r_3U_{{\bf p}}^{n-1}$, we write
\begin{align}
U_{{\bf p}}^{n+1}-U^*_{{\bf p}}=r_1\Delta t^{n+1} F^U_{{\bf p}}(\phi^{n+1}_{{\bf Q}},\phi^{*}_{{\bf p}},U^{n+1}_{{\bf Q}},U^*_{{\bf p}},\theta^{n+1}_{{\bf p}})
\label{U_time}
\end{align}
where
\begin{align}
F^U_{{\bf p}}(\phi_{{\bf Q}},\phi^{*}_{{\bf p}},U_{{\bf Q}},U^*_{{\bf p}},\theta_{{\bf p}})&\equiv \left. \left\{\frac{\nabla\cdot\left(D_c\frac{1-\phi}{2}\nabla U+{\b j}\right)+\frac{1}{2}[1+(1-k_E)U]\Par{\phi}{t}}{\left(\frac{1+k_E}{2}-\frac{1-k_E}{2}\phi\right)}\right\}\right|_{{\bf Q}}\nl
&\equiv \frac{D_c\frac{1-\phi_{\bf p}}{2}\nabla^2U|_{{\bf Q}}+\nabla(D_c\frac{1-\phi_{{\bf p}}}{2})|_{\bf Q-p}\cdot\nabla U|_{{\bf Q-p}}+\nabla\cdot{\bf j}|_{{\bf Q}}+\frac{1}{2}[1+(1-k_E)U_{{\bf p}}]\frac{\phi_{{\bf p}}-\phi^*_{{\bf p}}}{r_1\Delta t^{n+1}}}{\left(\frac{1+k_E}{2}-\frac{1-k_E}{2}\phi_{{\bf p}}\right)}.
\end{align}
The divergence of the anti-trapping current is given by
\begin{align}
\nabla\cdot{\bf j}|_{{\bf Q}}=-\frac{1}{2\sqrt{2}}[1+(1-k_E)]\nabla\cdot(U\dot\phi{\bf n})|_{{\bf Q}}
\end{align}
where
\begin{align}
\nabla\cdot(U\dot\phi{\bf n})|_{{\bf Q}}&=\dot\phi\nabla U\cdot {\bf n}+U\nabla\dot\phi\cdot{\bf n}+U\dot\phi\nabla\cdot{\bf n}\nl
&=\frac{\phi^{n+1}_{{\bf p}}-\phi^*_{{\bf p}}}{r_1\Delta t^{n+1}}\nabla U|_{{\bf Q-p}}\cdot {\bf n}|_{{\bf Q-p}}+U_{{\bf p}}\frac{\nabla\phi^{n+1}|_{\bf Q-p}-\nabla\phi^{*}|_{\bf Q-p}}{r_1\Delta t^{n+1}}\cdot{\bf n}|_{{\bf Q-p}}+U_{{\bf p}}\frac{\phi^{n+1}_{{\bf p}}-\phi^*_{{\bf p}}}{r_1\Delta t^{n+1}}\nabla\cdot{\bf n}|_{{\bf Q}}
\end{align}
and
\begin{align}
\nabla\cdot{\bf n}|_{{\bf Q}}\equiv\Par{}{x^a}\left(\frac{\phi_{,a}}{\sqrt{\phi_{,b}\phi_{,b}}}\right)|_{{\bf Q}}&=\frac{1}{\sqrt{\nabla\phi\cdot\nabla\phi}|_{\bf Q-p}}\left(\nabla^2\phi|_{{\bf Q}}-\left.\frac{\phi_{,a}\phi_{,ab}\phi_{,b}}{\nabla\phi\cdot\nabla\phi}\right|_{{\bf Q}}\right)\nl
&\equiv\frac{1}{\sqrt{\nabla\phi\cdot\nabla\phi}|_{\bf Q-p}}\left(\nabla^2\phi|_{{\bf Q}}-n_a|_{{\bf Q-p}}\phi_{,ab}|_{{\bf Q}}n_b|_{{\bf Q-p}}\right).
\end{align}
Finally, the heat equation with $\theta^*_{{\bf p}}\equiv r_2\theta_{{\bf p}}^{n}-r_3\theta_{{\bf p}}^{n-1}$we write
\begin{align}
\theta_{{\bf p}}^{n+1}-\theta^*_{{\bf p}}=r_1\Delta t^{n+1}  F^\theta_{{\bf p}}(\phi^{n+1}_{{\bf p}},\phi^{*}_{{\bf p}},\theta^{n+1}_{{\bf Q}},\theta^{*}_{{\bf p}})
\label{T_time}
\end{align}
where
\begin{align}
F^\theta_{{\bf p}}(\phi_{{\bf p}},\phi^{*}_{{\bf p}},\theta_{{\bf Q}},\theta^{*}_{{\bf p}})
&\equiv D_\theta \nabla^2\theta|_{\bf Q}+\ha\frac{\phi_{\bf p}-\phi^{*}_{\bf p}}{r_1\Delta t^{n+1}}
\end{align}

\subsection{Adaptive mesh and block tree structure}
\label{AMR}
The domain is first divided into a number of mesh  blocks each of which contains $N\times N\times N$ hexahedral cells, where we typically choose $N=8$.  We employ a domain of $800^3$, which is large enough for Lewis numbers of the order of $100$. We divide this domain into $4^3$ blocks, so that when N=8, each cell is of size $25^3$  and refer to this as level $1$. The adaptive mesh strategy then imposes a hierarchical sub-division of some or all of these blocks, and their descendants, based upon an oct tree structure. This subdivision aims to concentrate cells where gradients of the oct tree variables are highest and to ensure neighbour blocks differ by at most one level. The finest grid we work with is at level $7$, with a corresponding $\Delta x=25/2^7=0.1953125$. We find that the  minimum finest level necessary to obtain qualitatively reasonable results is level $5$, corresponding to $\Delta x=0.78125$. As noted previously, throughout this work we exploit cell-centred finite differences in our discretization, in which a single unknown is associated with the centre of each hexahedral cell for each of the dependent variables.

In order to discuss further the tree structure of the blocks we denote any block by its label, $i$ and its contents/properties, $B_i$:
\begin{align}
B(i)=[l,s,p,{\bf c},{\bf x}]
\label{B}
\end{align}
where $l\in[1,n]$ is the level, $s\in[1,8]$ is the sibling number (i.e. an index for which child of $p$ the block is), $p$ is the parent index, $c_i,i\in[1,8]$ are the 8 child indices, and ${\bf x}=[x,y,z]$ is the Cartesian position coordinates of the block origin. Any one of these properties can be accessed by the block number, $i$. Some examples: $p(i)$ is the block number of the parent of block $i$; ${\bf x}(p(i))$ is the position of the parent's origin; $c_{s(i)}(p(i))=i$ is an identity.
A natural position scheme for the child blocks is
\begin{align}
{\bf x}(c_1(i))&={\bf x}(i)+\ha\Delta x[-1,-1,-1]\nl
{\bf x}(c_2(i))&={\bf x}(i)+\ha\Delta x[1,-1,-1]\nl
{\bf x}(c_3(i))&={\bf x}(i)+\ha\Delta x[-1,1,-1]\nl
{\bf x}(c_4(i))&={\bf x}(i)+\ha\Delta x[1,1,-1]\nl
{\bf x}(c_5(i))&={\bf x}(i)+\ha\Delta x[-1,-1,1]\nl
{\bf x}(c_6(i))&={\bf x}(i)+\ha\Delta x[1,-1,1]\nl
{\bf x}(c_7(i))&={\bf x}(i)+\ha\Delta x[-1,1,1]\nl
{\bf x}(c_8(i))&={\bf x}(i)+\ha\Delta x[1,1,1]\nl,
\end{align}
where $\Delta x$ is the grid size associated with level $i$.
 A complete specification of all the blocks in the oct tree is then specified by the list:
\begin{align}
{\bf B}=\{B(i),i\in[1,B_N]\}
\label{List}
\end{align} 
where $B_N$ is the total block number. Moreover, a childless block, $i$, can be indicated by specifying, ${\bf c}(i)=0$ and so the oct tree also can be specified by a listing of just the leaf blocks
\begin{align}
{\bf B}=\{B(i),i\in[1,B_N]:{\bf c}(i)={\bf 0}\}.
\end{align}

There is no adaptive meshing within each block,which always contains $N\times N\times N$ cells and the adaptive strategy is further restricted by only allowing blocks at level $n$ adjacent to blocks of $n-1,n$ and $n+1$. Thus, even though a block may be flagged for coarsening, this (latter)  restriction often prevents this happening. Conversely, blocks flagged for refinement must, if necessary,  be accompanied by refinement on neighbouring blocks.  Blocks are flagged for refinement if, for any point, ${\bf p}$, in the block, the following criterion is satisfied:
\begin{align}
e\equiv\max \left\{e_\phi|\phi_{{\bf p}}-\phi_{{\bf p-q}}|,e_U|U_{{\bf p}}-U_{{\bf p-q}}|,e_T|T_{{\bf p}}-T_{{\bf p-q}}|\right\}> \eta,
\end{align}
where we use, for tolerance, $\eta\sim 1$ and
\begin{align}
|\phi_{{\bf p}}-\phi_{{\bf p-q}}|\equiv \sqrt{(\phi_{{\bf p}}-\phi_{{\bf p}-[1,0,0]})^2+(\phi_{{\bf p}}-\phi_{{\bf p}-[0,1,0]})^2+(\phi_{{\bf p}}-\phi_{{\bf p}-[0,0,1]})^2},
\end{align}
etc. Typically the weights, $e_\phi,e_U$ and $e_T$ are chosen to sum to unity. Sometimes it is convenient to set $e_U$ to zero to suppress unnecessary refinement within the solid. If $e<0.1\eta$, the block is flagged for derefinement.

Although each block is logically defined to be of dimension NxNxN the PARAMESH implementation actually allocates a block of dimension $(N+2G)\times(N+2G)\times(N+2G)$, where $G$ is the number of guard cells (sometimes referred to as ghost cells) around each block. When $G=1$, as used in this paper, the first and last cells in each dimension are guard cells - an update function may be called at any time in order to populate these guard cells with the corresponding values from the interior of each of the neigbouring blocks (with a separate treatment required to impose boundary conditions for blocks at the edge of the domain, as discussed previously). The application of a discrete stencil on any block requires access to neighborouring blocks via the guard cell nodes. When the neighbouring block is  coarser the guard cell of the  coarse block is found by a weighted average of the $8$ surrounding coarse nodes (some of which are in the parent block).  Using a tri-linear function of $x,y$ and $z$, gives the weightings, in order of nearest neighbours first
\begin{align}
{\bf w}=\left[\frac{27}{64},\frac{9}{64},\frac{9}{64},\frac{9}{64},\frac{3}{64},\frac{3}{64},\frac{3}{64},\frac{1}{64}\right].
\label{w}
\end{align} 
The process is known as prolongation. For example, the value of $\phi$ at a fine node is prolonged by
\begin{align}
\phi_\text{fine}=\sum_{i=\text{coarse cube centres}} w_i\phi_i.
\end{align} 
The reverse process, of finding a guard cell value for a coarse block when one or more neighbours is refined is known as restriction and is the simple average of the eight nearest, one level  finer, cell centres. Both operations, restriction and prolongation using Eq. \ref{w}, are  also required for multigrid as detailed in the next section.

\section{Solver method}
The discretisation above produces a system of nonlinear algebraic equations for $\phi^{n+1}_{{\bf p}},U^{n+1}_{{\bf p}}$ and $\theta^{n+1}_{{\bf p}}$ at each time step, $t^{n+1}$. In this section we describe the solution algorithm that is used to solve these systems, based upon a nonlinear multigrid (Full Approximation Scheme (FAS), \cite{Brandt1977}) approach. Initially we decribe the nonlinear smoother upon which the multigrid is built, before exploring how the multigrid solver combines this with the hierarchical mesh adaptivity introduced in the previous section. Finally, in subsection \ref{parallel}, we explain the key features of the parallel implementation, including the issues associated with parallel dynamic load-balancing.

\subsection{Nonlinear smoother}

The non-linear system of algebraic equations at the end of Sec. \ref{temporal} can be written using the generic vector notation ${\bf v}^{n+1}_{\bf p}\equiv [\phi^{n+1}_{\bf p},\theta^{n+1}_{\bf p},U^{n+1}_{\bf p}]$ by
\begin{align}
 {\bf A}_{\bf p}({\bf v}^{n+1}_{\bf Q})={\bf 0}
 \label{nonlinearsystem}
\end{align}
where
\begin{align}
{\bf A}_{\bf p}({\bf v}^{n+1}_{\bf Q})\equiv{\bf v}^*_{\bf p}-{\bf v}^{n+1}_{\bf p}+r_1\Delta t^{n+1}{\bf F}_{\bf p}({\bf v}^{n+1}_{\bf Q})
\end{align}
for each node, ${\bf p}$, in the grid. Recall, that the appearance of ${\bf Q}$ indicates coupling between points ${\bf p}$ and neighbouring nodes, and the BDF2 notation ${\bf v}^*_{\bf p}$ in combination with $r_1$ are defined in subsection \ref{temporal}. Using an iteration method, with ${\bf v}^{n+1}_{\bf p}$ approximated by ${\bf v}^{n+1,m}_{\bf p}$, we define the defect
\begin{align}
{\bf d}^{n+1,m}_{\bf p}=-{\bf A}_{\bf p}({\bf v}^{n+1,m}_{\bf Q}).
\end{align}
 The pointwise  Newton update for this iteration is given by
\begin{eqnarray}
{\bf v}_{\bf p}^{n+1,m+1}= {\bf v}_{\bf p}^{n+1,m}-\omega \tilde{\bf d}_{\bf p}^{n+1,m},
\end{eqnarray}
where $\tilde{\b d}^{n+1,m}_{\bf p}$ is found by solving the $3\times 3$ system
\begin{align}
\b J^{n+1,m}_{\bf p}\tilde{\b d}^{n+1,m}_{\bf p}\equiv{\bf d}^{n+1,m}_{\bf p}
\end{align}
with the $3\times 3$ Jacobian matrix defined by
\begin{align}
\b J^{n+1,m}_{\bf p}\equiv\Par{{\bf d}^{n+1,m}_{\bf p}}{{\bf v}^{n+1,m}_{\bf p}}.
\end{align}
In practice we typically select, $\omega\approx 0.9$ and find that off diagonal terms of ${\bf J}^{n+1,m}_{\bf p}$ are not essential to obtain a convergent iteration. 

The precise form of the entries of  ${\bf J}^{n+1,m}_{\bf p}$ may be deduced from the above. However, we illustrate this in detail for one of the diagonal terms for the sake of clarity. Denoting the diagonal entries of ${\bf J}^{n+1,m}_{\bf p}$ by $[J^{\phi,n+1,m}_{\bf p},J^{\theta,n+1,m}_{\bf p},J^{U,n+1,m}_{\bf p}]$, using Eqs. \ref{phi_time} and \ref{phirhs}, and treating terms not including $\phi_{\bf p}$ as constant we find
\begin{align}
J^{\phi,n+1,m}_{\bf p}\equiv \Par{d^\phi_{\bf p}({\bf v}_{\bf Q}^{n+1,m})}{\phi_{\bf p}}=\left[1+r_1\Delta t^{n+1}\tfrac{1}{3}\frac{g_{11}+g_{22}+g_{33}}{\tau(c,\phi)A^2}\frac{128}{30(\Delta x)^2}\right]_{\bf p},
\end{align}
where we note that the contribution from the central node to $\nabla^2\phi$ is
\begin{align}
\Par{\nabla^2\phi|_{\bf Q}}{\phi_{\bf p}}=-\frac{128}{30(\Delta x)^2}.
\end{align}
The new updated solution for $\phi$ at time $t^{n+1}$ at iteration $m$ and point ${\bf p}$ is given by
\begin{align}
\phi^{n+1,m+1}_{\bf p}= \phi_i^{n+1,m}-\omega \frac{d^{\phi,n+1,m}_{\bf p}}{J^{\phi,n+1,m}_{\bf p}}.
\end{align}
The term $J^{\theta,n+1,m}_{\bf p} =1+r_1\Delta t^{n+1} D_\theta\frac{128}{30(\Delta x)^2}$ and, though the term $ J^{U,n+1,m}_{\bf p}$ has many components, the linearity of $F^U_{\bf p}$ in $U_{\bf p}$ leads to straightforward updates for the  $U$ components, n.b. gradients involving $U|_{\bf Q-p}$ are treated as constant.

The above describes a point wise non-linear Jacobi smoother. The Jacobi approach lends itself to parallel implementation since the computation at each point may be completed using neighbouring values from the previous iteration only. Consequently, only one ghost cell update is required prior to each sweep through the mesh. This keeps inter processor communication to a minimum. Using the stencils described above it is possible to complete the updates using just one layer of ghost cells. Hence if a block size of $8\times 8\times 8$ (say) is used in PARAMESH then a $10\times 10\times 10$ block is actually allocated to accommodate the ghost layer of each block edge.

Having derived the point-wise Jacobi smoother, in the following subsection we show how, this may be used as part of a non-linear geometric multigrid solver  combined with local mesh adaptivity. Discussion of the parallel implementation is postponed until subsection \ref{parallel}.
\begin{figure}
\fbox{
\begin{minipage}{17cm}
{\bf FAS algorithm to solve} ${\bf A(v)=0} $

\begin{enumerate}
\item $h=$ finest grid (top level)
\item ${\bf v}^{n+1,0}_{{\bf p}(h)}\leftarrow{\bf v}^{n}_{{\bf p}(h)}$: set initial guess equal to value at last time step\\
\item ${\bf v}^{n+1,\nu+1}_{{\bf p}(h)}\leftarrow ${\bf V-cycle}$({\bf v}^{n+1,\nu}_{{\bf p}(h)},{\bf 0},h)  $: application of V-cycle, $\nu=0,1,...$ until convergence
\end{enumerate}
{\bf Recursive Function V-cycle}$({\bf v}^{0}_{{\bf p}(h)},{\bf f}_{{\bf p}(h)},h)\rightarrow{\bf v}^{m+1}_{{\bf p}(h)}   $ (solves ${\bf A(v)=f}$)
\begin{enumerate}
\item  ${\bf v}^{m+1}_{{\bf p}(h)}\leftarrow {\bf S}({\bf v}^{m}_{{\bf p}(h)},{\bf f}_{{\bf p}(h)},h)\equiv\left\{
\begin{array}{l}
 {\bf d}^{m}_{{\bf p}(h)}\leftarrow {\bf f}_{{\bf p}(h)}-{\bf A}({\bf v}^{m}_{{\bf Q}(h)})\\
{\bf v}^{m+1}_{{\bf p}(h)}\leftarrow {\bf v}^{m}_{{\bf p}(h)}-\omega\left[\frac{{ d}^{\phi,m}_{{\bf p}(h)}}{{ J}^{\phi,m}_{{\bf p}(h)}},\frac{{ d}^{\theta,m}_{{\bf p}(h)}}{{J}^{\theta,m}_{{\bf p}(h)}},\frac{{d}^{U,m}_{{\bf p}(h)}}{{ J}^{U,m}_{{\bf p}(h)}}\right]
\end{array}
\right.,\quad m=0,..., M-1
$\\
 (pre) smooth $M$ times, (typically we use $M\sim 4$) 
  \item     ${\bf d}^{m+1}_{{\bf p}(h)}\leftarrow {\bf f}_{{\bf p}(h)}-{\bf A}_{\bf p}({\bf v}^{m+1}_{{\bf Q}(h)})$\\ 
  ${\bf v}^{0}_{{\bf p}(2h)}\leftarrow {\bf I}^h_{2h}({\bf v}^{m+1}_{{\bf p}(h)})$\\
   ${\bf d}^{0}_{{\bf p}(2h)}\leftarrow {\bf I}^h_{2h}({\bf d}^{m+1}_{{\bf p}(h)})+{\bf A}({\bf v}^{0}_{{\bf p}(2h)})$: restriction
\item {\bf if} not coarsest level\\
${}\quad{\bf v}^{m+1}_{{\bf p}(2h)}\leftarrow$ { \bf V-cycle}$({\bf v}^{0}_{{\bf p}(2h)},{\bf d}^{0}_{{\bf p}(2h)} ,2h) $.\\
{\bf else} (bottom level)\\
 ${}\quad{\bf v}^{m+1}_{{\bf p}(2h)}\leftarrow {\bf S}({\bf v}^{m}_{{\bf p}(2h)},{\bf d}^{0}_{{\bf p}(2h)} ,2h),\quad m=0,...,N$ (typically we use $N\sim 4$)\\
 {\bf end if}
\item ${\bf d}^{0}_{{\bf p}(2h)}\leftarrow {\bf v}^{m+1}_{{\bf p}(2h)}-{\bf v}^{0}_{{\bf p}(2h)}$: compute the correction (now on upward part of v-cycle)
\item ${\bf v}^{0}_{{\bf p}(h)}\leftarrow {\bf v}^{m+1}_{{\bf p}(h)}+{\bf I}^{2h}_h({\bf d}^{0}_{{\bf p}(2h)})$:  prolong and correct fine grid solution
\item ${\bf v}^{m+1}_{{\bf p}(h)}\leftarrow {\bf S}({\bf v}^{m}_{{\bf p}(h)},{\bf 0},h),\quad m=0,...,M_\text{post}-1$: (post) smooth with this new value $M_\text{post}\sim 4$ times
\end{enumerate}

\end{minipage}
}
\caption{The FAS algorithm}
\label{FAS}
\end{figure}
\subsection{Nonlinear Multigrid}
\label{Nonlinear}
Although the Jacobi smoother described above gives a convergent iteration for the system Eq. \ref{nonlinearsystem} (for sufficiently small $\Delta t$ and good initial guess), the convergence is far too slow to be of any practical use. Fortunately, however, the iteration also satisfies a smoothing property which means that it damps out the highest frequency components of the error (defect) far more quickly than the rest. This makes it ideally suited for use as part of a nonlinear multigrid scheme. In this work we make use of the Full Approximation Scheme (FAS) of Brandt \cite{Brandt1977}. We denote the value of variables at point, ${\bf p}$, time $t^{n+1}$, iteration, $m$, and level/grid size, $h$ by ${\bf v}^{n+1,m}_{{\bf p}(h)}$ and coarser level ${\bf v}^{n+1,m}_{{\bf p}(2h)}$.  The restriction operation, ${\bf I}^h_{2h}({\bf v}_{{\bf p}(h)})$, is the assignment to  ${\bf v}_{{\bf p}(2h)}$  of the simple average of the 8 surrounding  nodes, ${\bf v}_{{\bf p}(h)}$. Prolongation,  ${\bf I}^{2h}_{h}{\bf v}_{{\bf p}(2h)}$, is an assignment to ${\bf v}_{{\bf p}(h)} $ given by applying  Eq. \ref{w}  to give a weighted average of the values at the 8 nearest coarse nodes ${\bf v}_{{\bf p}(2h)} $. In solving Eq. \ref{nonlinearsystem}, FAS computes a  defect from the restricted defect and variables to give a modified ${\bf A(v)= f} $ on these lower levels. See  Fig. \ref{FAS}, where we detail FAS for our notation.

\begin{figure}
\begin{center}
\includegraphics[width=12cm]{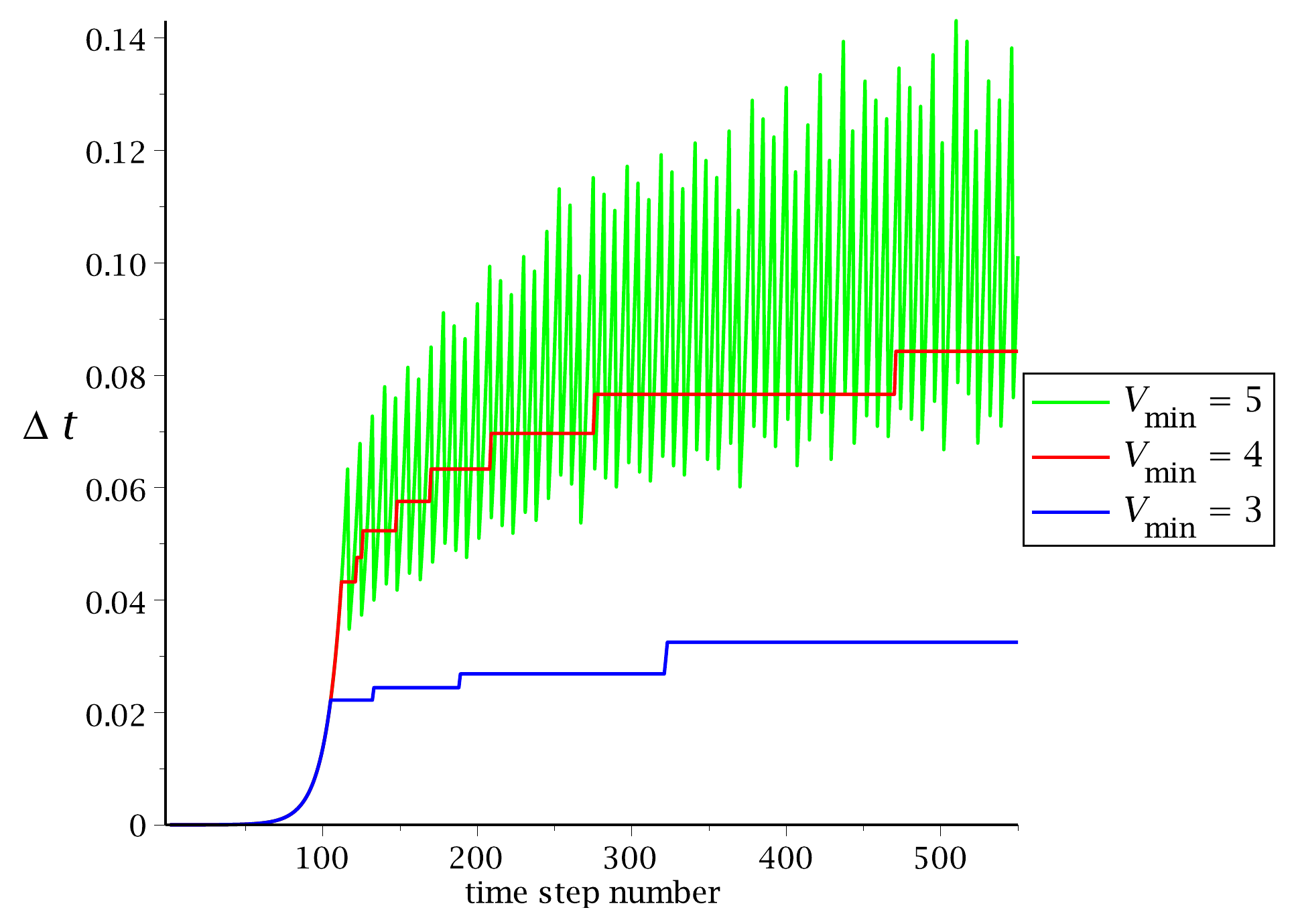}
\end{center}
\caption{An illustration, for $Le=40,\Delta=0.525$ of the evolution of the time step, $\Delta t$, from $10^{-6}$ by V-cycle control.}
\label{timestep}
\end{figure}
Note that in our work local mesh adaptivity is an essential feature. This has been described in Sec. \ref{AMR}, where PARAMESH block data structures are used as nodes of an oct-tree. In this work we also use the oct-tree as part of the geometric multigrid solver by developing an implementation of the Multi Level Adaptive Technique (MLAT) (see \cite{Brandt1977}). MLAT allows us to use the Jacobi smoother on each block without modification, provided the prolongation and restriction operators are adapted to deal with guard cells of the interface between too differing levels of refinement. Briefly, the smoother is only applied in the regions of the domain that contains fine level blocks, but the coarse grid correction takes place on all parts of the domain that contains the coarse level blocks (though the modified right hand side associated with the FAS scheme only contributes to those coarse grid regions that have fine grids on them).

For the results presented in the following sections we are primarily interested in obtaining solutions at large times.
Hence we choose a time-stepping strategy with this in mind, based upon the number of nonlinear V-cycles that are
required to achieve convergence (for an alternative strategy, which uses a local error estimate to control $\Delta t$,
see \cite{Rosam2007}). The principle is simple: if the nonlinear multigrid converges easily at a give time step then
increase $\Delta t$, whereas if it converges slowly (or fails to converge) then decrease $\Delta t$ (repeating the
time step in the case of failure). Convergence is deemed to have occurred when the infinity norm of the defect
(possibly weighted differently for each dependent variable), $d$, satisfies $d < d_{\mbox{max}}$, for a user-defined
stopping parameter $d_{\mbox{max}}$. If this is not satisfied in $V_{\mbox{fail}}$ V-cycles then $\Delta t$ is halved
and the step is retaken. If convergence occurs in $V_{\mbox{min}}$ V-cycles or less then $\Delta t$ is increased by
$10 \%$ however if convergence occurs in more than $V_{\mbox{max}}$ V-cycles then $\Delta t$ is halved for the
next step. Figure \ref{timestep} shows a typical evolution of the time step size for three different choices of
$V_{\mbox{min}}$, based upon initial $\Delta t = 10^{-6}$, $d_{\mbox{max}} = 10^{-10}$ and $V_{\mbox{max}}=10$.
Note that although the oscillation in $\Delta t$ is aesthetically undesirable it has no adverse effect on the solution
quality nor (so long as $V_{\mbox{max}} < V_{\mbox{fail}}$) the overall efficiency.

\subsection{Parallel implementation}
\label{parallel}
 Our implementation requires communication between blocks both on the same level (to apply the smoothing steps) and between levels (for prolongation and restriction). For parallel processing, clearly, communication between cores needs to be kept to a minimum, but an important secondary consideration is that each core has as near as possible equal load. For a uniform mesh an allocation of each core to an equal volume of the computational domain results in an obvious fair division of labour. On the other hand, for  an adaptive mesh, such an approach fails since the loading between cores will differ enormously.

Given the label, $i$,  of each block in Eq. \ref{B}, we present the Morton ordering, $M(i)\in[1,B_N]$, of an adaptive mesh
\begin{align}
M(i)=\left\{
\begin{array}{ll}
1&l=1,s=1 \text{ (bottom level)}\\
M(p(i))+1,& s=1,l>1 \text{(one plus parent's label)}\\
M(s(i-1))+1,& s>2, \text{ and } l= \text{  leaf  level }\\
M(c_8(i))+1,&\text{ otherwise, i.e.  } s=2
\end{array}
\right.,
\end{align}
  where `leaf level' refers  an unrefined level (parent without child) and includes the finest level. The blocks may then be put into a Morton ordered list:
\begin{align}
{\bf M}(\textbf{B})&\equiv \{B_i:j=M(i),j\in[1,B_N]\}\nl
&\equiv \{...,B_{M(i)},B_{M(j)},...:M(i)<M(j) \}
\label{Morton}
\end{align}
We write the form given in second line of Eq. \ref{Morton} to highlight the difference between Morton ordering and the alternative ordering we adopt below, Eq. \ref{Level} and \ref{Li}.

Load balancing for $B_N$ blocks on processors $p_i,i=1,2,...,c_N$ follows the same order with approximately $B_N/c_N$ blocks per core. The resulting ordering is well known to exhibit parallel inefficiency for non-uniform meshes. This is because Morton ordering on adaptive meshes leaves the majority of the top level blocks on a small fraction of cores and  since multigrid advances from the top to bottom level and back sequentially, the majority of the cores will be idle at the top level. In three dimensions this problem is acute because there is a factor of 8 between levels. 

We adopt the following ordering, which may be termed Morton-{\it Level} ordering.
\begin{align}
{\bf L}(\textbf{B})&\equiv\{B_i:j=L(i),j\in[1,B_N]\},\nl
&\equiv\{{\bf L}_k({\bf B}),k\in[1,n]\},
\label{Level}
\end{align}
where the subsets, ${\bf L}_k({\bf B})$ on each level, $k$ are defined
\begin{align}
{\bf L}_k({\bf B})&\equiv\{...,B_{M(i)},B_{M(j)},...:M(i)<M(j),l(i)=l(j)=k \}
\label{Li}
\end{align}

In summary, we implement a depth first traversal of the blocks and then, using this numbering, divide the work load at each level in turn between all processors. For a uniform mesh this strategy results in a near optimum allocation to cores. For adaptive meshes the communication on any level is also optimum, but communication between level is compromised.

\section{Computational Results}

This section provides a selection of computational results that are designed to validate and assess our proposed solution algorithm. Since this is the first attempt to produce three dimensional results for the solification of a non-isothermal alloy using a realistic interface width we have no external simulations against which to validate our code. Hence the approach taken here has been to firstly validate a two-dimensional restriction of our implementation against our own two-dimensional solver, implemented completely independently and described in \cite{Rosam2007}. These tests show that we are indeed able to reproduce results from \cite{Rosam2009}, (e.g. the tip radius and velocities are in excellent agreement) even though the two code bases are completely independent, e.g. in \cite{Rosam2009} forth order accurate stencils were used. 

Furthermore, we have also successfully validated two 3-d simplifications of our implementation. In \cite{Green2011} we consider a thermal-only restriction (i.e. a pure metal, so no concentration equation present), where we show quantitative agreement
with results obtained using the 3-d, explicit, thermal-only approach of \cite{Jeong2001,Provatas1999}. Similarly, in \cite{Goodyer2012} we show quantitative
agreement between an isothermal version of our solver (i.e. no temperature equation present) and another 3-d isothermal solidification code described in \cite{Jeong2001}.To validate the 3-D non-isothermal simulations we now rely on having  mesh convergence and multigrid performance.

\begin{figure}
\begin{center}
\includegraphics[width=15cm]{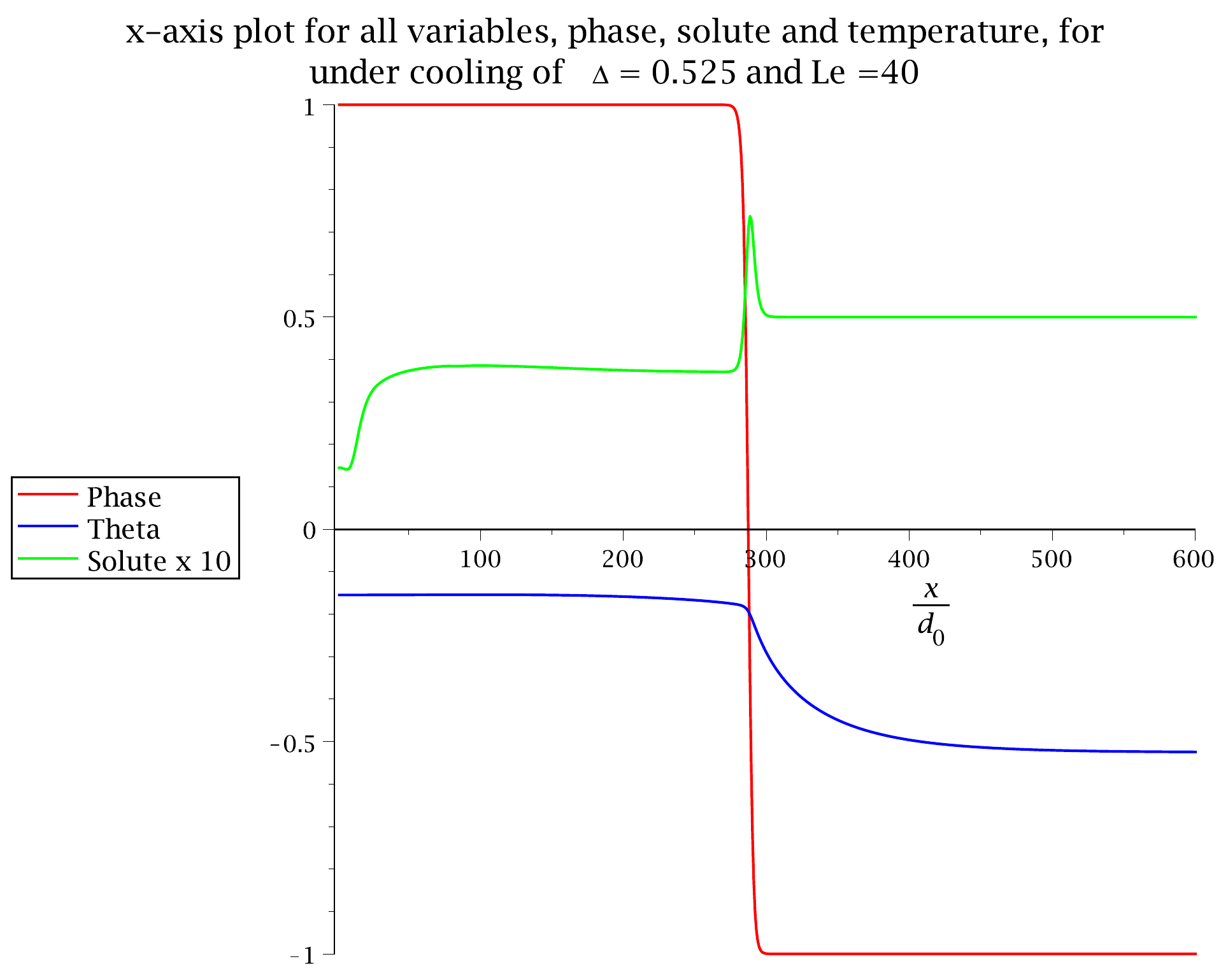}
\end{center}
\caption{This plot shows the variables: phase $\phi$, solute concentration $c$, and dimensionless temperature $\theta$. At Lewis number of 40 and under cooling of $0.525$ the temperature diffusion zone extends to about 300 in a  domain size  of $800^3$}
\label{var}
\end{figure}
\begin{figure}
\begin{center}
\includegraphics[width=15cm]{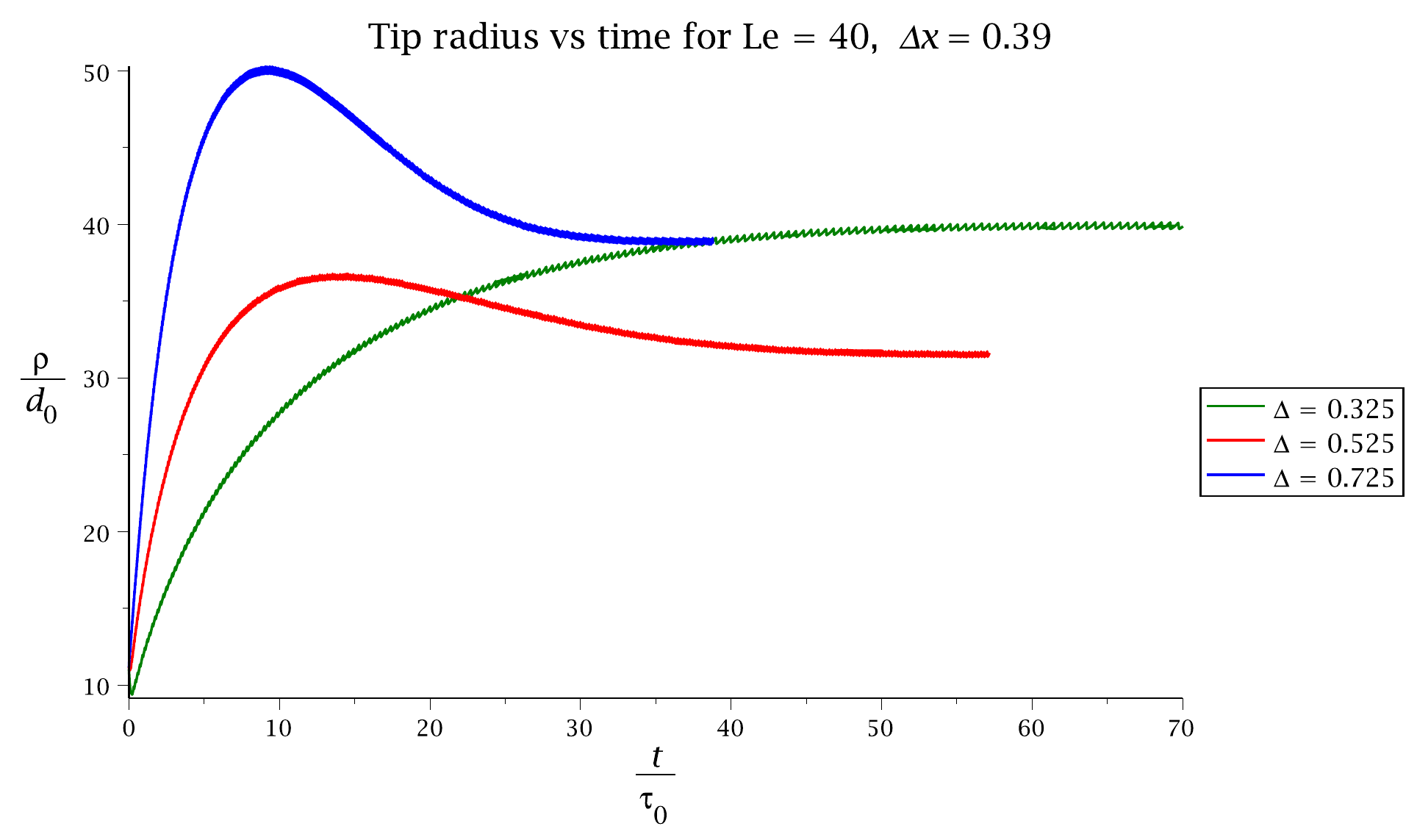}
\end{center}
\caption{Tip radius at $dx=0.39$ and $Le=40$ for a range of under cooling $\Delta$.  Even though the radius at $\Delta=0.325$ takes longer to reach steady state, this simulation is faster than the others due to greater stability resulting in fewer, larger time steps.}
\label{m1}
\end{figure}
\begin{figure}
\begin{center}
\includegraphics[width=15cm]{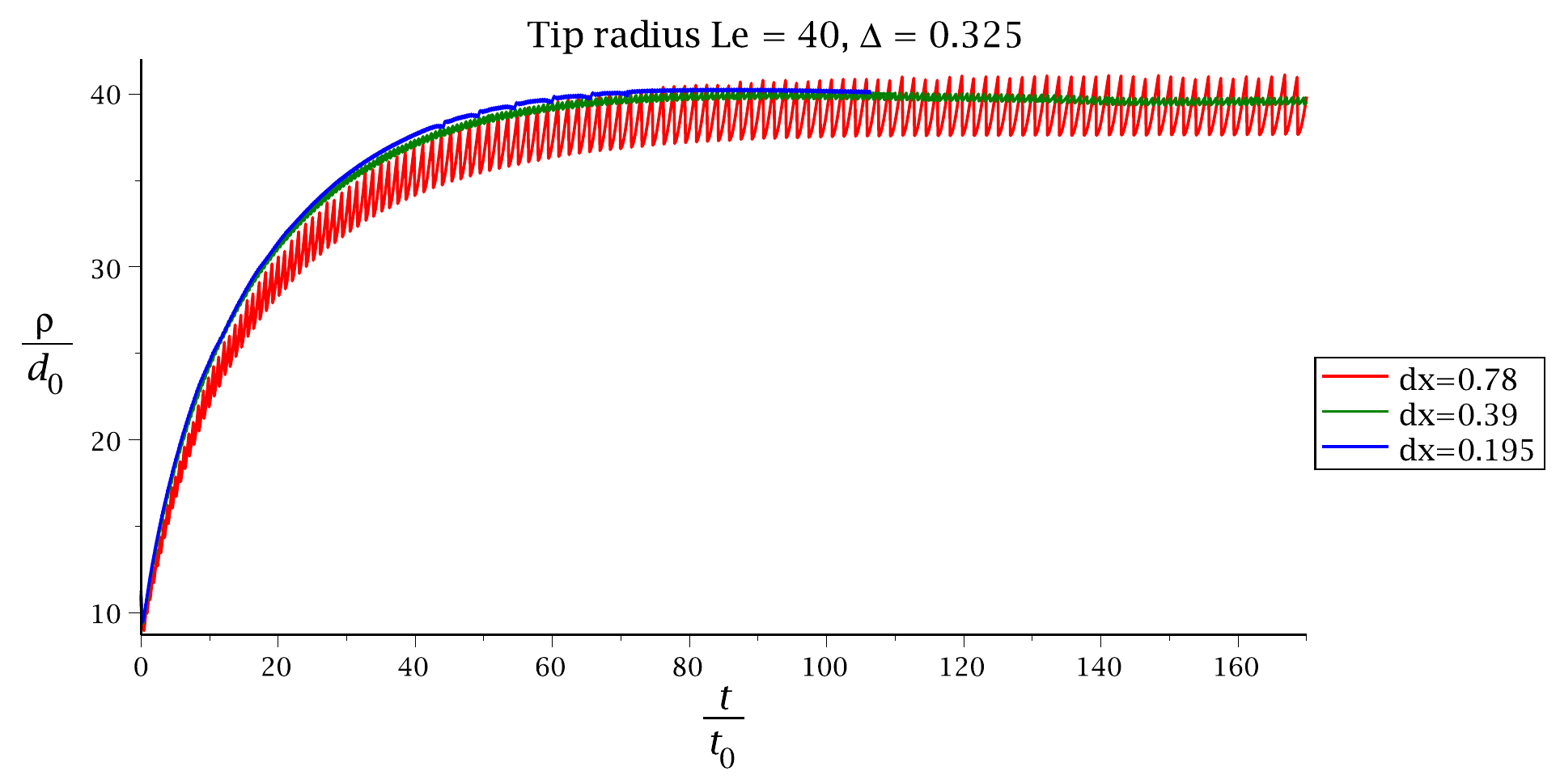}
\end{center}
\caption{Convergence test on tip radius with grid sizes $dx \in[0.78,0.39,0.195]$. The plot shows, for $\Delta =0.325$, the full transient behaviour of the tip radius for all grid resolutions. In this case the results for all three grids is in good agreement. Higher under cooling makes the coarser grid less reliable and even $dx=0.39$ becomes less reliable.}
\label{m3}
\end{figure}

\begin{table}
\begin{center}
\begin{tabular}{|c|c|c|c|c|}
\hline
\multicolumn{2}{|c|}{Parameters} &\multicolumn{3}{|c|}{Tip radius}\\
\hline
$Le$ &$\Delta$ &$\Delta x=0.78$ &$\Delta x=0.39$ &$\Delta x=0.195$ \\
\hline
$40$&$0.325 $&$ 39.0\pm 2$ &$ 39.7\pm0.3$ &$ 39.7\pm0.1$\\
$40$&$0.525 $&$29.7\pm 2 $&$ 31.5\pm0.3$ &$ 31.8\pm0.1$ \\
$100$&$0.325 $&$40.3\pm 2$ &$ 39.8\pm0.3$&$ 39.8\pm0.1$\\
\hline
\end{tabular}
\end{center}
\caption{Table of tip radius results for two under-coolings and two Lewis numbers. For this particular selection of parameters there is reasonable agreement even on the coarser mesh $\Delta x=0.78$. But the slight discrepancy shown for the higher under-cooling is symptomatic of the necessity for a finer mesh,  $\Delta x > 0.39$, in general. }
\label{Results}
\end{table}

The remainder of this section is divided into two subsections. The first of these considers the mesh convergence of our implementation, showing that large time solutions obtained on a sequence of finer levels of maximum refinement do indeed appear to converge to particular dendrite geometries, as tested for a selection of parameter values. The second subsection considers the numerical performance of the solver. In particular it is shown that optimal performance is achieved, whereby the time required to complete a time step grows almost linearly with the total number of degrees of freedom. The capability improvements associated with the distributed memory parallel implementation are also discussed.
\subsection{Mesh Convergence}
In order to gain further confidence in our computational approach we have undertaken a number of mesh convergence tests, in which we considered computational simulations in which the maximum level of mesh refinement is systematically increased. In order to appreciate the need for very fine grids at the phase interface Fig. \ref{var} illustrates a cross section along the x-axis of a typical solution. This corresponds to the same parameter values as used to compute the dendrite illustrated in Fig. \ref{m9}. It is clear that the phase variable, $\phi$, changes $+1$ (bulk solid) to $-1$ (bulk melt) over a very small distance, similarly the solute concentration varies rapidly both at, and immediately ahead of, the interface. The temperature variable, $\theta$, decays much more smoothly however -- though a large domain is required to ensure that the boundary effect does not contaminate the solution. In addition to the interface width a further feature of significant interest is the geometry of the dendrite tip. Fig. \ref{m1} shows the computed evolution of the tip radius for three different undercoolings ($0.325,0.525$ and $0.725$) at $Le=40$. The parameter, $d_0=5\sqrt{2}/(8\lambda)$ is the (non-dimensional) capillary length as a function of the interface width. Our value for $\lambda=2$ gives, $d_0=.44$ indicating that the interface width in our simulation is of the order of the physical width. We use as our characteristic time scale, $t_0=0.80(\tilde d_0^2\lambda^3)/\tilde D_c$ where $\tilde d_0$ and $\tilde D_c$ are the physically dimensioned capillary length and solute diffusivity coefficient respectively. Note that the results for Fig. \ref{m1} were computed using a fine mesh spacing of $\Delta x= 0.39$.  It is an important question to ask if this is sufficiently fine for the solution to be insensitive to further mesh refinement.

Fig. \ref{m3} shows the computed tip radius as a function of time for three different maximum refinement levels ($\Delta x=0.78,0.39 $ and $0.195$) for case $Le=40$ and $\Delta=0.325$. The tip radius involves estimating a second derivative of $\phi$ in the region where $\phi=0$. This is undertaken by estimating the radius on the x-axis at $\phi=0$ using the phase field:
\begin{align}
r=\left.\frac{\phi_{,x}}{\phi_{,uu}}\right|_{\phi(x,y,z)=0,y=0,z=0}
\label{radius}
\end{align}
where $\Par{}{u}=\frac{1}{\sqrt 2}\left(\Par{}{y}+\Par{}{z}\right)$, $\phi_{,x}\equiv \Par{\phi}{x}$ and $\phi_{,uu}\equiv\Par{{}^2\phi}{u^2}$ (as will become clear shortly, the $u$ direction is more convenient than the $y$ or $z$ directions). Expression Eq. \ref{radius} comes from the definition
\begin{align}
r=1/\kappa
\end{align}
where the curvature, $\kappa$, is defined in the normalised direction ${\bf u}$ to be
\begin{align}
\kappa = {\bf u}\cdot\nabla{\bf n}\cdot{\bf u}.
\end{align}
On the x-axis, ignoring the $z$ direction, and with a directional derivative in the $y$ direction, we find, using $\phi$ to compute the normal ${\bf n}\equiv\nabla\phi/{|\nabla\phi|}$, that
\begin{align}
\kappa = \Par{ n_2}{y}=\Par{}{y}\frac{\phi_{,y}}{\sqrt{\phi_{,x}^2+\phi_{,y}^2}}=\frac{\phi_{,yy}}{\phi_{,x}}
\end{align}
where, by symmetry $\phi_{,xy}|_{y=0}=\phi_{,y}|_{y=0}=0$ on the axis. Again, by symmetry, this relation holds for any normalised parameter  and so
\begin{align}
\kappa = \frac{\phi_{,yy}}{\phi_{,x}}= \frac{\phi_{,zz}}{\phi_{,x}}= \frac{\phi_{,uu}}{\phi_{,x}}
\end{align}
In practice, of course, the value $\phi=0$ lies between two successive nodes on the x-axis, $i$ and $i+1$. Furthermore, the x-axis lies, by definition, on $y=z=0$ where there are no grid points. We compute the derivatives only using the points
\begin{align}
[i,2,2],[i+1,2,2],[i,3,3],[i+1,3,3]
\end{align}
which the relates  to the physical points
\begin{align}
[i,j,k]\rightarrow [O_x,0,0]+\Delta x[i-3/2,j-3/2,k-3/2]
\end{align}
where $O_x$ is the x coordinate of the block origin. Thus, we know by symmetry, that the values at these points can be equated to the image nodes (which we do not use explicitly)
\begin{align}
\phi_{[i,1,1]} &=\phi_{[i,2,2]},\nl
\quad \phi_{[i,0,0]}&=\phi_{[i,3,3]}.
\end{align}
We compute the radius of curvature using the direction $u$ by
\begin{align}
r=\left.\left(\frac{\phi_x}{\phi_{uu}}+\frac{\phi}{\phi_x}\right)\right|_{[i,2,2]}
\end{align}
where the last term is a correction to compensate for, in general, $\phi\ne 0$ at the point $ [i,2,2]$. This is discetised by
\begin{align}
\phi_x &=\frac{\phi_{i+1,2,2}-\phi_{i-1,2,2}}{2\Delta x},\nl
\phi_{uu} &=\frac{-\phi_{i,2,2}+\phi_{i,3,3}}{2(\Delta x)^2}
\end{align}
It is the nature of this approximation that leads to the oscillatory results that are observed on the least fine simulation ($\Delta x=0.78$). Nevertheless it is clear that the results for $\Delta x=0.39$ and $\Delta x=0.195$ are almost indistinguishable at the scale used here and so we have a good degree of confidence that a converged solution is obtained by $\Delta x=0.39$. Similar computations of the large-time tip radius, on different levels of grid refinement, have been undertaken for two other cases, as shown in Tab. \ref{Results}. Whilst the convergence is not so mature in every case, the evidence that results at $\Delta x=0.39$  are of sufficient accuracy to be of quantitative validity  is very strong.
\begin{figure}
\begin{center}
\includegraphics[width=15cm]{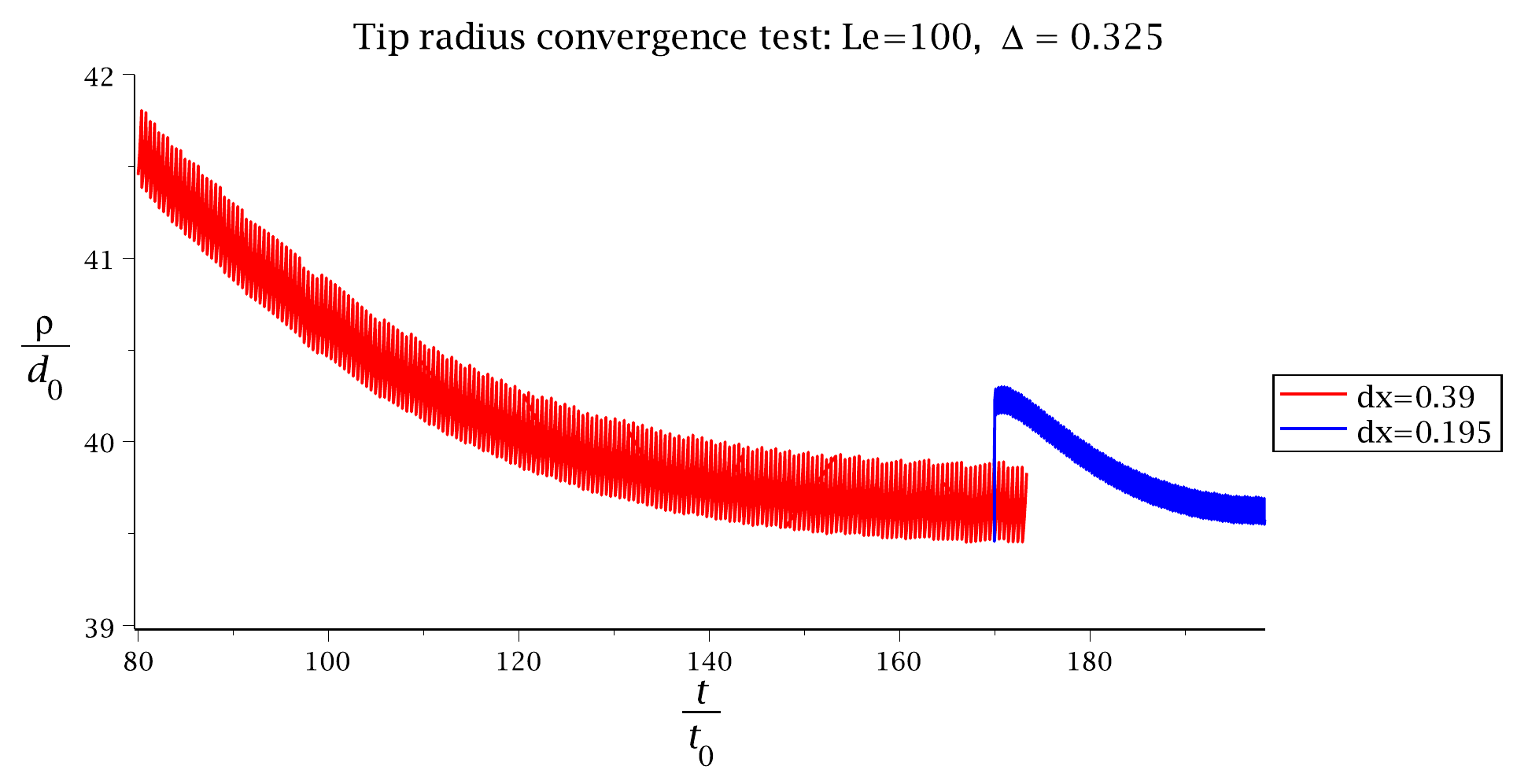}
\end{center}
\caption{Convergence test on tip radius with grid sizes $dx \in[0.39,0.195]$. A checkpoint file at $dx=0.39$ is used as an initial condition for a $dx=0.195$ to test convergence. The restart recovers from an initial transient before settling to a value very similar to the steady state at the coarser, $dx=0.39$ run.}
\label{m2b}
\end{figure}

\begin{figure}
\begin{center}
\includegraphics[width=15cm]{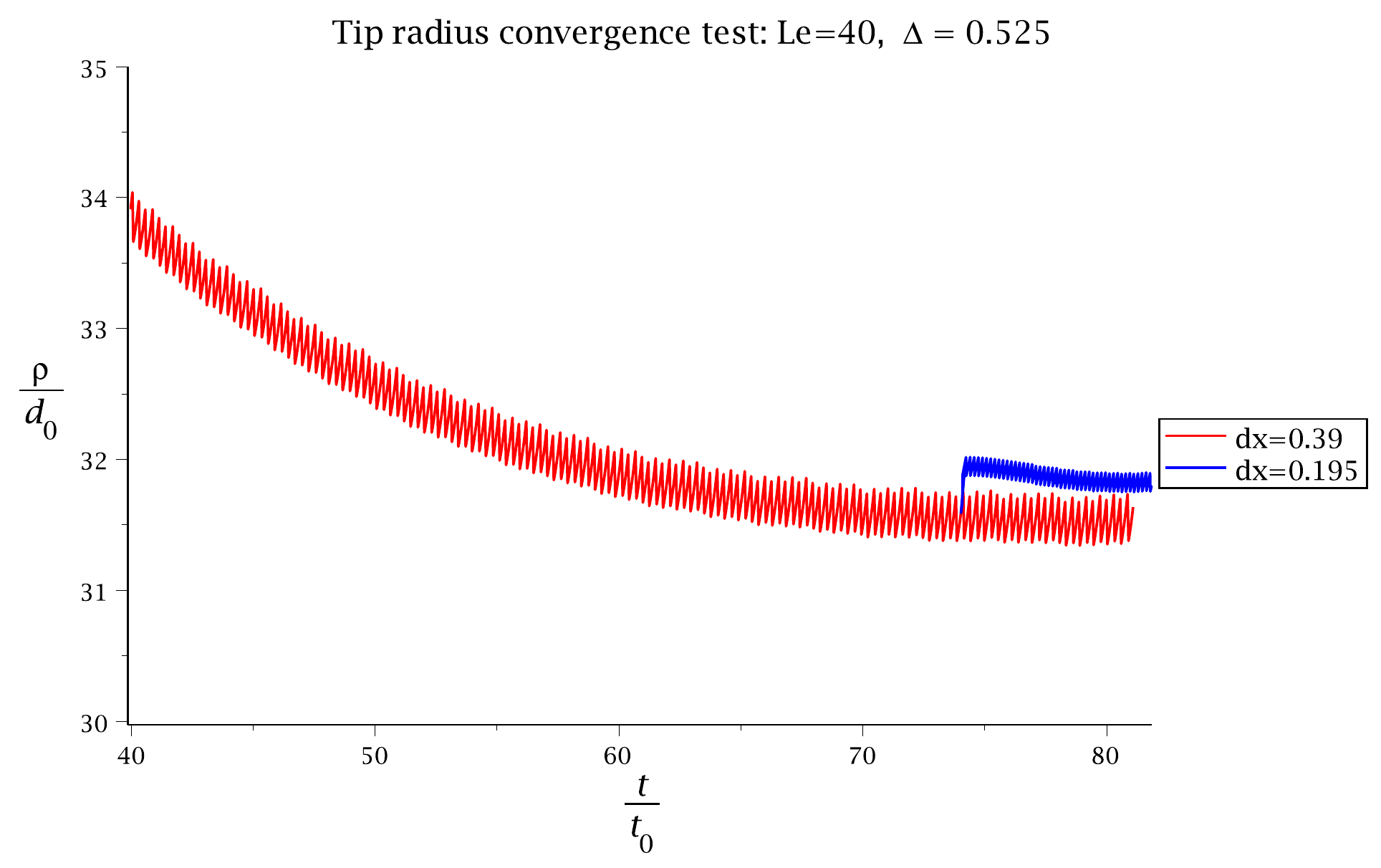}
\end{center}
\caption{Convergence test on tip radius with grid sizes $dx \in[0.39,0.195]$. A checkpoint file at $dx=0.39$ is used as an initial condition for a $dx=0.195$ to test convergence.}
\label{convtest525}
\end{figure}

\begin{figure}
\begin{center}
\includegraphics[width=15cm]{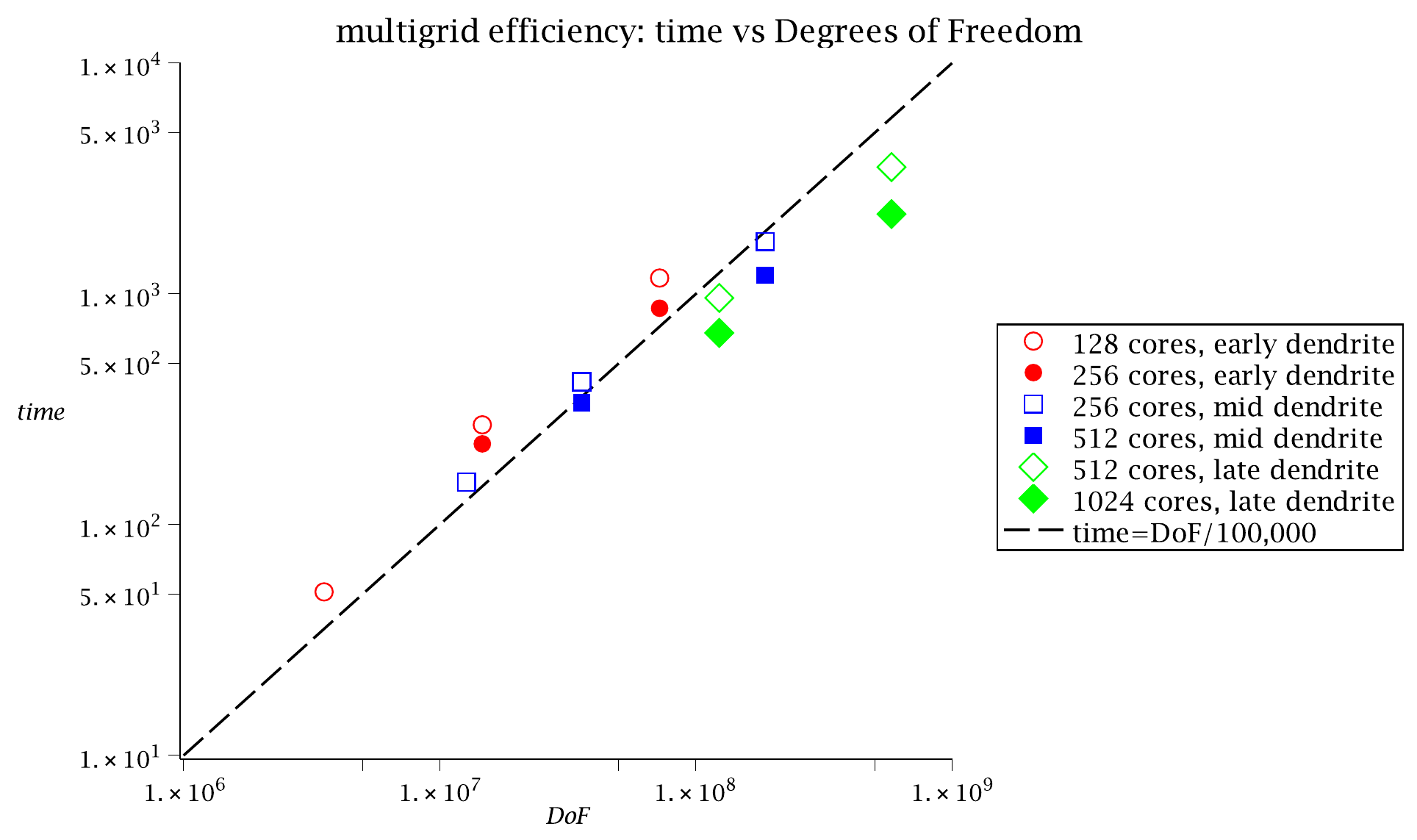}
\end{center}
\caption{Multigrid efficiency. A demonstration of the linearity of solve time with the number of degrees of freedom. In all 6 cases the corresponding points fit well (for a single time step) to the line of slope 1.}
\label{s2}
\end{figure}
\begin{figure}
\begin{center}
\includegraphics[width=15cm]{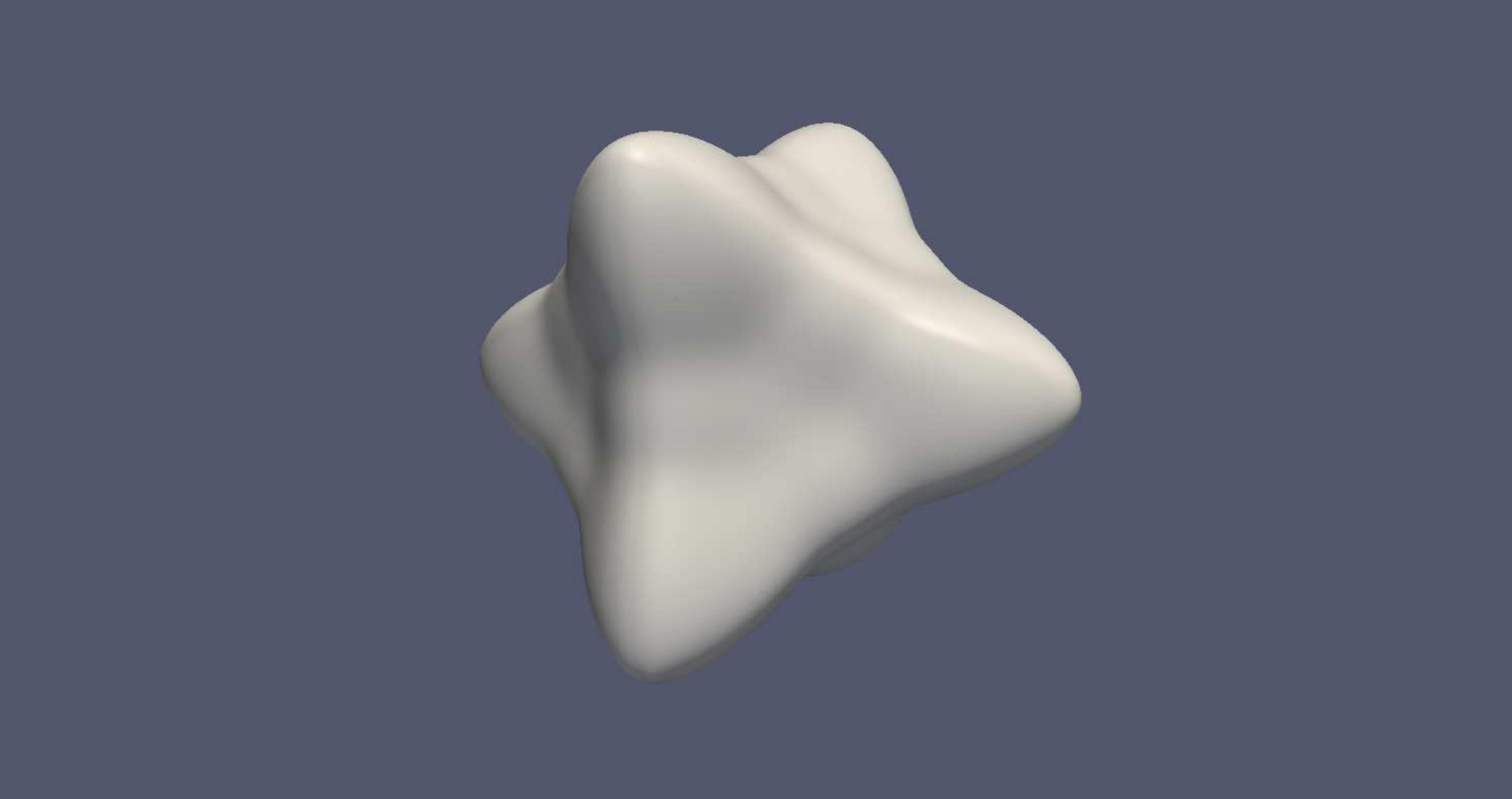}
\end{center}
\caption{Dendrite image: $Le=40,\theta=-0.525, t=102,  \de x=0.78$. This simulation took 12 hours on a 12 core machine.}
\label{EarlyDendrite}
\end{figure}
\begin{figure}
\begin{center}
\includegraphics[width=15cm]{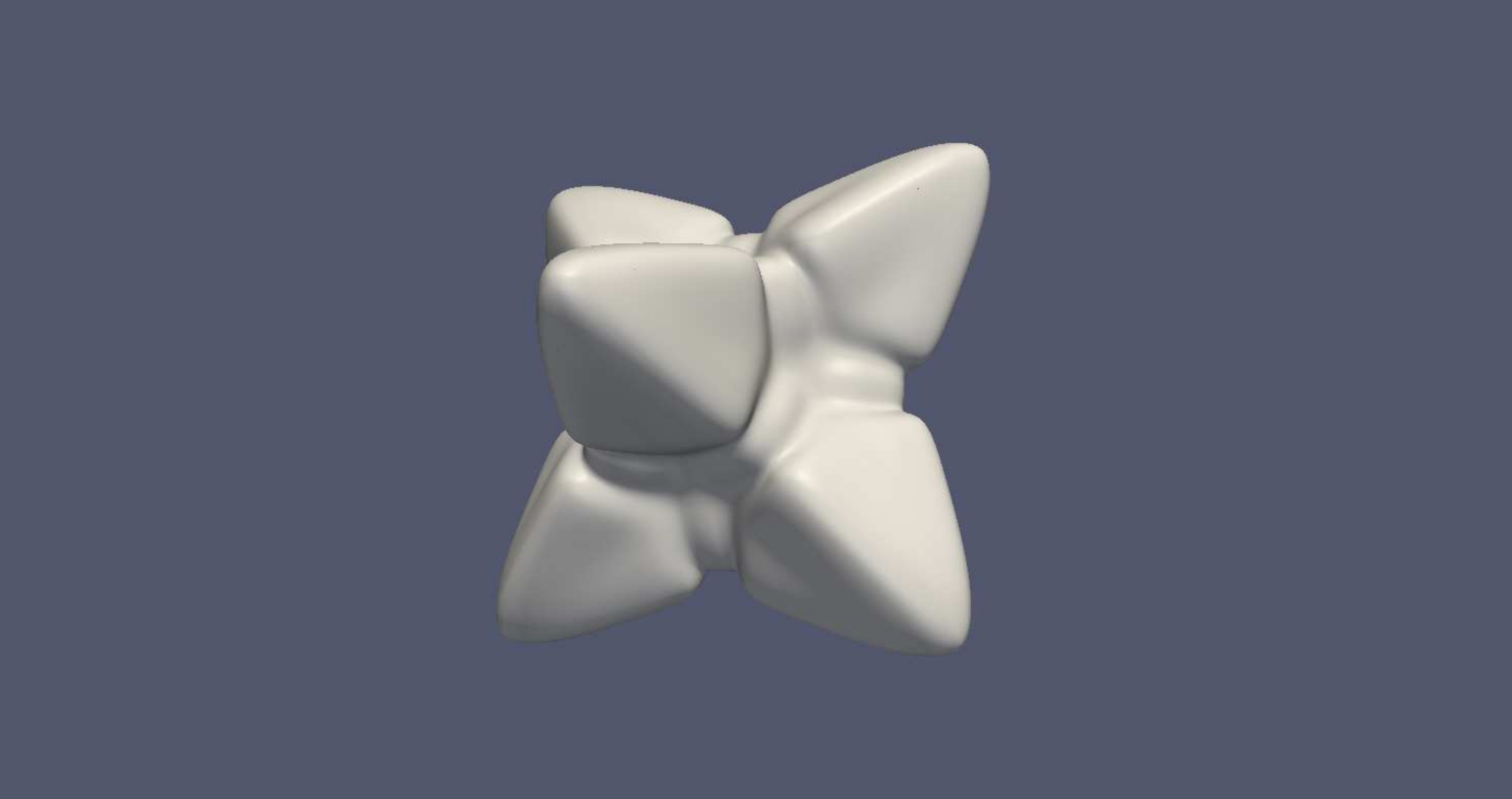}
\end{center}
\caption{Dendrite image: $Le=40,\theta=-0.525, t=186,  \Delta x=0.78$. This simulation is of a dendrite with the coarsest maximum refinement and took about 40 hours to simulate on a 12 core machine. The last 10\% of the run (a  time interval of $t=18$)  took about 10 hours.}
\label{MidDendrite}
\end{figure}

\begin{figure}
\begin{center}
\includegraphics[width=15cm]{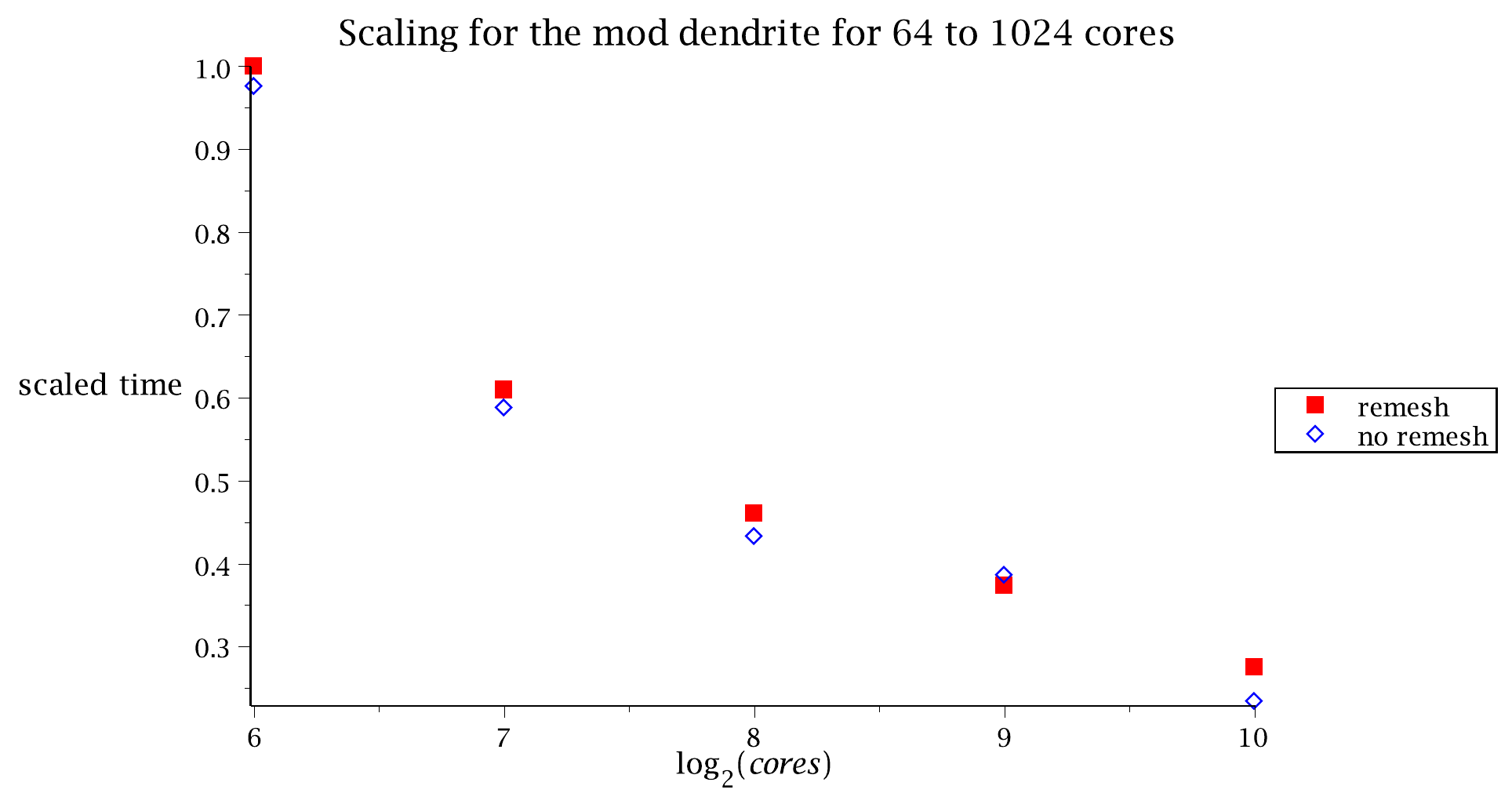}
\end{center}
\caption{Plot of the relative wall-clock time to undertake 10 time steps at the mid dendrite stage (see Fig. \ref{MidDendrite}), for mesh $dx=0.39$,  using different numbers of cores (64 to 1024). }
\label{s1}
\end{figure}
\begin{figure}
\begin{center}
\includegraphics[width=15cm]{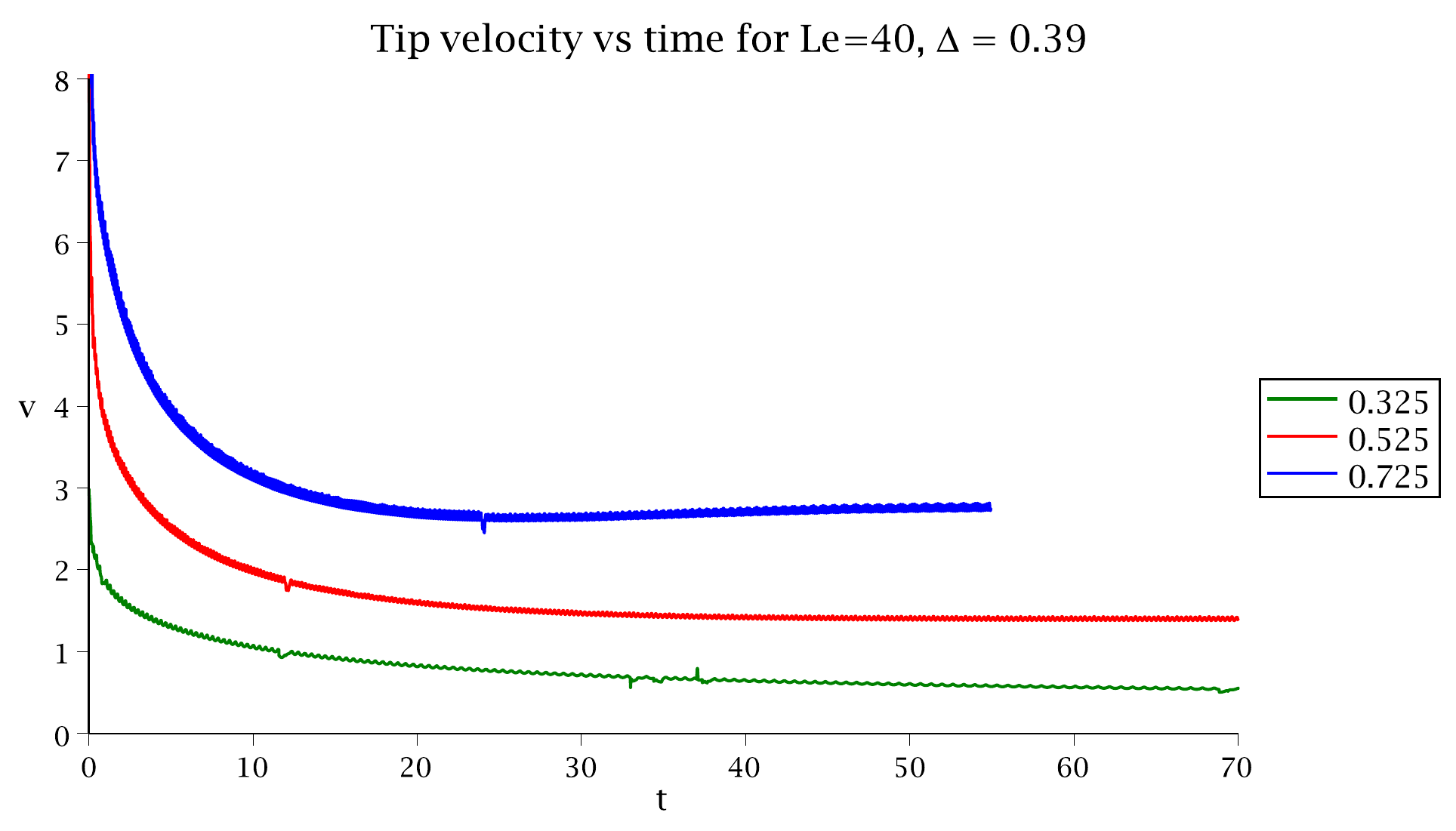}
\end{center}
\caption{This is a companion plot to the tip radius plot \ref{m1}:  The tip velocity at $dx=0.39$ and $Le=40$ for a range of under cooling $\Delta=0.325,0.525,0.725$. Though the tip radius has reached steady state  in each of these cases the tip velocity for the highest undercooling cases is still increasing.}
\label{mm}
\end{figure}

Note that the $\Delta x=0.195$ results for the other cases shown in Tab. \ref{Results} were not obtained by undertaking complete runs at this maximum refinement level. Instead, the large-time simulation computed using a maximum level of $\Delta x=0.39$ was restarted with a maximum level of $\Delta x=0.195$ and executed until a steady state tip radius was reached. This was tested for the ($Le=40,\Delta=0.325$) case, and shown to give identical results. The approach is illustrated for  for $Le=100,\Delta=0.325$ in Fig \ref{m2b} and $Le=40,\Delta=0.525$ in Fig. \ref{convtest525}.

\subsection{Numerical Performance}
Having demonstrated the mesh convergence of our proposed technique in the previous subsection, we now consider the computational performance of the implementation used.

The most important feature of any successful multigrid implementation is that it should enable solutions of systems of algebraic equations to be obtained in a run time that is close to $O(N)$ as $N\rightarrow\infty$ where $N$ is the number of degrees of freedom. In this simulation we begin with a small solid seed at the origin and this grows (under that right parameter conditions) in time. As it grows the region of maximum mesh refinement gets larger and larger, as the isosurface, $\phi=0$ has ever increasing area. This causes $N$ to increase in time. Hence, an excellent test of our solver is whether the computed time to take each time step only grows in proportion to $N$. Unfortunately this test is harder to undertake than initially  might  be imagined since, as $N$ increases the amount of memory required to compute a time step also increases (linearly). As described in Sec 4 we deal with this through a distributed memory parallel implementation. We start the execution, with a small seed, using a modest number of cores (16 or 32 say) and increase these as the execution progresses. Furthermore, as discussed in the previous subsection, we also wish to consider different choices for the maximum level of mesh refinement, which also impacts on the number of computational nodes to be used and therefore the  total memory requirement.

Fig \ref{s2} shows a selection of timings for a set of computations for a single, representative, test case ($Le=40,\Delta =0.525$) using different cores. The vertical axis shows the execution time for a single implicit time step and the horizontal axis shows the number of degrees of freedom (both on log scales). Timings are taken just as the dendrite is starting to form (early dendrite see Fig. \ref{EarlyDendrite}), part way through its formation (mid dendrite Fig. \ref{MidDendrite}), and once the dendrite is clearly formed (late dendrite Fig. \ref{m9}). Timings are also taken using different maximum refinement levels (either 2 or 3 per case). It is very clear from Fig. \ref{s2} that multigrid performance is achieved throughout these different stages of the evolution of the dendrite and at different maximum refinement levels. This can be seen from the excellent proximity  of the points to the time line of slope one ($t=DoF/100,000$) on the log-log scale.

Note that our use of distributed memory parallel computing throughout this work has been aimed primarily at providing the capacity to solve large systems (up to and beyond a billion degrees of freedom per implicit time step) in a computationally efficient manner. It is clear from Fig. \ref{s1} that the strong parallel scalability of our implementation is not optimal. Nevertheless it is apparent that, as well as providing additional memory capacity to allow larger problems to be tackled, out parallel implementation also continues to improve the speed of the execution each time each time the core count is increased so long as the number of degrees of freedom is sufficiently large.
\section{Conclusions}
We have presented, in detail, the mathematical model and  methods employed to simulate, for the first time, a three dimensional, fully coupled thermal-solute-phase field model for  dendritic growth. This was achieved through the coupling of multiple numerical techniques from compact discrete finite difference stencils, AMR, MLAT and parallel execution.

For moderate Lewis numbers we have been able to obtain results at sufficient grid resolution, which we have confidence can provide quantitative accuracy in 3-d for the first time. This break through into three dimensional fully coupled thermal simulation is of importance, since heat generation at the growing two dimensional surface is very much an integral part of the natural physical process and two dimensional results therefore have little quantitative value. Our next goal is to undertake a systematic simulation of the effect of increasing Lewis number.

We end with the observation that, though, tip radius is fully converged, the tip velocity is only near steady state. We give the companion plot to Fig. \ref{m1} for the tip velocities for the same parameters in Fig. \ref{mm}.

\section{Acknowledgements}

This research was funded by EPSRC grant number EP/H048685.  We are also grateful for the use of  the HECToR UK National Supercomputing Service.

\bibliography{petermetric.bib}
\bibliographystyle{unsrt}

\end{document}